\documentclass{emulateapj}
\usepackage{psfig}

\def\msol{\hbox{$\rm\thinspace M_{\odot}$}}
\def\etal{{\it et al.\thinspace}}
\def\eg{{\it e.g.\ }}

\def\fig{Figure }

\def\p3m{P${}^3$M}
\def\ap3m{AP${}^3$M}
\def\-{{\em{---}}}

\def\msun{{M_\odot}}

\newcommand{\be}{\begin{equation}}
\newcommand{\ba}{\begin{eqnarray}}
\newcommand{\ee}{\end{equation}}
\newcommand{\ea}{\end{eqnarray}}  
\received{}   
\accepted{}   

\lefthead{Thacker, Scannapieco, \& Couchman}
\righthead{Quasars: What turns them off?}

\begin{document}

\def\lesssim{\mathrel{\hbox{\rlap{\hbox{\lower4pt\hbox{$\sim$}}}\hbox{$<$}}}}
\def\gtrsim{\mathrel{\hbox{\rlap{\hbox{\lower4pt\hbox{$\sim$}}}\hbox{$>$}}}}
\def\ApJ{{\em Astrophys.\ J.\ }}
\def\AJ{{\em Astron.\ J.\ }}
\def\MNRAS{{\em Mon.\ Not.\ R.\ Astron.\ Soc.\ }}
\def\CIV{\hbox{C$\scriptstyle\rm IV\ $}}\

\title{Quasars: What turns them off?}

\author{Robert J. Thacker\altaffilmark{1,2}, Evan Scannapieco\altaffilmark{3}, and
H. M. P. Couchman\altaffilmark{4}}
\altaffiltext{1}{Department of Physics, Engineering Physics and Astronomy,
Queen's University, Kingston, Ontario, K7L 3N6,
Canada.}
\altaffiltext{2}{CITA National Fellow}
\altaffiltext{3}{Kavli Institute for Theoretical Physics,
 Kohn Hall, University of California, Santa Barbara, CA 93106}
\altaffiltext{4}{Department of Physics and Astronomy,
McMaster University, 1280 Main St.\ West, Hamilton, Ontario, L8S 4M1,
Canada.}

\begin{abstract} 

While the high-redshift quasar luminosity function closely parallels
the hierarchical growth of dark matter halos, at lower redshifts
quasars exhibit an anti-hierarchical turn-off, which moves from the
most luminous objects to the faintest.    We explore the idea that
this may arise from self-regulating feedback, caused by quasar
outflows.  Using a detailed hydrodynamic simulation we calculate the
luminosity function of quasars down to a redshift of $z=1$ in a large,
cosmologically representative volume.  Outflows are included
explicitly by tracking halo mergers and driving shocks into the
surrounding intergalactic medium, with an energy output equal to a fixed  
5\% fraction
of the bolometric luminosity.   Our results are in excellent agreement
with measurements of the spatial distribution of quasars on both small
and large scales, and we detect an intriguing excess of galaxy-quasar
		   pairs at very short separations.
Our results also reproduce an anti-hierarchical turnoff in the
quasar luminosity function,
however, this falls short of that observed as well as
that predicted by analogous semi-analytic models.  The difference can
be traced to the treatment of heating of gas within galaxies and the
presence of in-shock cooling. Calculations of the mass fraction  of
gas above the critical entropy show a strong redshift dependence
with close to 20\% of the baryons being above this limit at $z=1$.
Volume fractions show an even stronger trend with redshift and some
sensitivity to resolution due to the tendency of high entropy gas to
occupy low density regions. The simulated galaxy cluster $L_X-T$
relationship is close to that observed for $z\approx 1$ clusters,
but the simulated galaxy groups at  $z=1$ are significantly perturbed by
quasar outflows, suggesting that measurements of X-ray emission in 
high-redshift groups could well be a ``smoking  gun'' for the AGN heating
hypothesis.

\end{abstract}

\section{Introduction}

In the low-redshift universe, active galactic nuclei (AGN)  are  not
very active.   While at high redshifts,  the quasar luminosity
function increases with time, since $z \approx 2$  the number density
of optically-selected AGN has been dropping dramatically (Schmidt \&
Green 1983; Boyle \etal 1988 Koo \& Kron 1988; Pei 1995; Boyle et
al. 2000; Fan et al. 2001).  Deep X-ray surveys have shown that  this
downturn occurs anti-hierarchically, such that the spatial density of
AGN with higher X-ray luminosities  peaks earlier than that of
lower-luminosity AGN (Steffen \etal 2003; Ueda \etal 2003).
Complementary emission line studies suggest that this trend 
is driven
by a decrease in the characteristic mass of  actively growing black
holes (Heckman \etal 2004), and is likely to closely parallel the
formation history of early-type galaxies (\eg Granato \etal 2001, 2004).
Furthermore, optical and near-infrared observations indicate
that the largest galaxies were already in place by z $\approx 2$, 
while smaller ones continued to form stars at much lower redshifts 
(Pozzetti \etal 2003; Fontana et al. 2004; Glazebrook et al. 2004; 
van Dokkum et al. 2004; Treu et al. 2005), and a similar trend
is observed in the morphological evolution of galaxies 
(Bundy, Ellis, \& Conselice 2005).

Yet, despite these observations, such widespread galaxy
``downsizing'' (Cowie \etal 1996) was 
unexpected.  The $\Lambda$ Cold Dark Matter
($\Lambda$CDM) model, while in spectacular agreement with observations
(\eg Spergel \etal 2003), is a hierarchical theory,  in which
gravitationally-bound structures grow by accretion and merging.
Superimposed on this distribution is the baryonic component, which
falls into the dark-matter potential wells, shock heats, and must
radiate this energy away before forming stars (Rees \& Ostriker 1977;
Silk 1977).  The larger the structure, the longer it takes to cool,
and thus galaxy evolution should be even more hierarchical than
structure formation.

Recently, several theoretical studies have shown that the missing
element in this picture could be kinetic feedback from AGN.  As bulge
and  black hole masses are closely related (Gebhardt \etal 2000;
Merritt \& Ferrarese 2001; see also King 2003, 2005), such outflows 
would 
have 
the largest
impact on the largest forming elliptical galaxies, 
suppressing their formation first (Scannapieco \& Oh 2004,
hereafter SO04;  Binney 2004; Di Matteo \etal 2005; Croton  \etal
2006).  This would also help to explain the X-ray
luminosity-temperature relationship observed in the intracluster
medium (ICM) in galaxy clusters.  If nongravitational heating were
unimportant, the gas density distribution would be self-similar,
resulting in $L_X \propto T^2$  (Kaiser  1986), but instead the
observed slope steepens considerably  for low-temperature clusters
(\eg David \etal 1993; Arnaud \& Evrard 1999; Helsdon \& Ponman 2000).
Furthermore the $100$ keV cm$^{2}$ level of preheating necessary to
explain this discrepancy (Cavaliere \etal 1998; Kravtsov \& Yepes
2000;  Wu \etal 2000;  Babul \etal 2002)  is not arbitrary.  Rather it
corresponds to the threshold value for cooling within the age  of the
universe (Voit \& Bryan 2001; Oh \& Benson 2003).

Taken together, these observations are strongly suggestive of a model
in which the properties of the ICM, the formation history of
elliptical galaxies, and the evolution of the quasar luminosity
function are all set by self-regulating AGN feedback.  In fact, S004
have shown that the addition of  AGN outflows into the
semi-analytic model developed by  Wyithe \& Loeb (2002; 2003) can
reproduce  the drop in the AGN luminosity at low redshifts, as heating
this gas slows the accretion of matter onto the supermassive black
hole.   The central AGN engine is then starved of fuel and  a strong
suppression at the bright end of the luminosity function occurs.  The main drawback of
the model, however, is that precise matching of observational  results
requires fine tuning of the model parameters, such as black hole  mass
and outflow efficiency.

Several other recent numerical and analytic investigations have
sharpened our understanding of various aspects of this process.
Binney (2004) conducted an analytic study of the impact of AGN on
inhibiting gas cooling in large galaxies.  Di Matteo \etal (2005)
carried out Smoothed Particle Hydrodynamic (SPH) simulations of AGN
outflows in individual galaxy mergers and studied the role of
feedback in determining the  colors of elliptical galaxies and
establishing the relationship between stellar velocity
dispersion and black hole mass.  This suite of simulations was later
related analytically to the global evolution of  quasars and
elliptical galaxies in series of papers by Hopkins \etal (2005a, 2005b,
2005c, 2006).  Levine \& Gnedin (2005) combined cosmological
simulations with an analytic model  to constrain the filling factor of
AGN outflows as a function of  redshift.  Scannapieco, Silk, \& Bouwens
(2005) emphasized the role that quasars may play in the downsizing
of the star-forming galaxy population.  Levine \& Gnedin (2006)
studied the impact of AGN outflows on the matter power spectrum.
Menci \etal (2006) used a semianalytical model to
study the role of AGN feedback on the color
distribution of galaxies from $z=0$ to $z=4.$
The importance of the nature of gas accretion in determining
the effectiveness of feedback processes has recently been
discussed in Dekel \& Birnboim (2006) and Cattaneo \etal (2006).
Finally, Croton \etal (2006) combined
semi-analytic models with the dark-matter evolution taken from
the Millennium  Simulation (Springel \etal 2005) to study the impact
of a more temporally-extended model of
AGN feedback on the bright end of the galaxy luminosity function.

In this paper we undertake the first detailed hydrodynamic simulations 
of quasar outflows in a general cosmological context.  Adopting a burst 
model that associates AGN with merger events and a global outflow
model that is based on our simulations of  high-redshift 
starbursts (Scannapieco, Thacker, \& Davis 2001, 
hereafter STD01), we are able track in detail the impact that
quasars have both on their own formation, and on the properties
of galaxy clusters and the intergalactic medium (IGM).

The structure of this work is as follows.  In \S 2 we describe our
overall numerical approach, method for quasar identification, and
implementation of outflows.  In \S 3 we compare our simulation results with 
measures of quasar clustering and the observed luminosity function
in the optical and the X-ray bands. In \S 4 we study AGN
feedback in a more global context, examining its impact on the 
properties of the intergalactic and
intracluster media. In \S5 we present a discussion and conclusion.

\section{Simulations of Quasar Formation}

\subsection {Overall Numerical Model}

Motivated by measurements of the cosmic microwave background, the number
abundance of galaxy clusters, and high-redshift supernova distance
estimates (\eg  Spergel et al. 2003; Vianna \& Liddle 1996; Riess \etal 
1998; Perlmutter \etal 1999), we focus our attention on a Cold Dark Matter
cosmological model with parameters $h=0.7$, $\Omega_0$ = 0.3,
$\Omega_\Lambda$ = 0.7, $\Omega_b = 0.046$, $\sigma_8 = 0.9$, and
$n=1$, where $h$ is the Hubble constant in units of 100 km s$^{-1}$
Mpc$^{-1}$, $\Omega_0$, $\Omega_\Lambda$, and $\Omega_b$ are the total
matter, vacuum, and baryonic densities in units of the critical
density, $\sigma_8^2$ is the variance of linear fluctuations on the $8
h^{-1}{\rm Mpc}$ scale, and $n$ is the ``tilt'' of the primordial
power spectrum. The  Eisenstein \& Hu (1999) transfer function is used
throughout.

As in our earlier work (STD01), simulations were conducted with a
parallel OpenMP based implementation of the ``HYDRA'' code (Thacker \&
Couchman 2006)  that uses the Adaptive Particle-Particle,
Particle-Mesh algorithm (Couchman 1991) to calculate gravitational
forces, and the SPH method (Lucy 1977; Gingold \& Monaghan 1977) to
calculate gas forces.  Gas densities and energies are calculated using
the standard SPH smoothing kernel method (for exact details see
STD01),  with the kernel tuned  to smooth over 52 particles; and
radiative cooling is calculated using standard tables (Sutherland \&
Dopita 1993). We have  kept the metallicity constant at $Z=0.05$, to
mimic a moderate level of enrichment in the galaxy  formation
process. However, this is an underestimate of the  intracluster
metallicity, $Z=0.3$, which seems to be an approximately universal
value to intermediate redshifts (Tozzi \etal 2003). Lastly,  because
the epoch of  reionization is poorly known, and because we are
primarily focusing our  attention on mass scales greater than
$10^{10}$\msol, we do not include a fiducial photoionization
background in the simulation.

We also do not include the so called ``$\nabla h$'' terms (Nelson \&
Papaloizou 1994, Serna \etal 1996, Springel \& Hernquist 2002) in our 
implementation of
SPH. This is potentially a  significant concern in this investigation
as we will examine gas  entropy.  However, tests on an expanding
spherical shell  problem (see STD01) using 1000 particles and no
artificial viscosity or cooling, show  entropy conservation to be
accurate at the 6\% level, while the combined  gravitational and
hydrodynamic energy error is around 1.5\%.  These findings are in
broad agreement with the  discussions presented in Hernquist (1993)
and Springel \& Hernquist  (2002), where a small entropy conservation
error was always accompanied  by a larger energy error when
integrating the evolution of the entropic  function in the absence of
$\nabla h$ terms. While a  lower error in the  entropy is desirable,
for the present phenomenological  investigation and given the gains we 
get
from using a code without $\nabla  h$ terms, we consider this error
acceptable.

We simulated a number of different box sizes and particle numbers to
quickly assess the accuracy and numerical resolution  dependencies in
our model. A single large simulation  was then run for statistical
purposes, allowing us to probe the bright  end of the luminosity
function. The specifics of each  simulation, including box size and
resolution are given in Table 1. We  note that attempting  to simulate
the  formation of the very brightest end of the luminosity function
with  sufficient resolution to track smaller mergers is  a difficult
task, due to the scarcity of these objects. This is the  fundamental
motivation behind  our progressing to a simulation with $2\times640^3$ 
particles.

\begin{table*}
\begin{center}
\caption{Simulation parameters}
\begin{tabular}{lccccc}
\hline\hline
\vspace{1mm}
Run  & N & Box Size & Mass Resolution & Softening Length & Steps (z=1)
\\
& & (Mpc) & (M${}_\odot$) & (kpc) &
\\
\hline \vspace{1mm}
0400 & $2\times40^3$ & $35 \;{h}^{-1}$ & $1.2\times10^{10}$ & $24\;
h^{-1}$ & 1340 
\\
0800 & $2\times80^3$ & $35 \;{h}^{-1}$ & $1.5\times10^{9}$ & $12\;
h^{-1}$& 3678 \\
1600  & $2\times160^3$ & $35 \;{h}^{-1}$ &$1.9\times10^{8}$ & $6\;
h^{-1}$ & 5420 
\\
3200  & $2\times320^3$ & $35 \;{h}^{-1}$ & $2.3\times10^{7}$ & $3 \;
h^{-1}$ & 7381 (z=2.5) \\
1020 & $2\times640^3$ & $146 \;{h}^{-1}$ & $2.2\times10^{8}$  & $9\;
h^{-1}$ &10420 (z=1.2) 
\\                                                                                
\hline
\smallskip
\end{tabular}
\end{center}
\end{table*}

\subsection {Identification of Quasars}

A secondary motivation of this paper is to compare the simulation
results directly with semi-analytic predictions. We have therefore
taken  the outflow model of SO04 and adapted it to our simulation as
closely as  possible, although in some cases, which we highlight, it
was either not possible for us to match this model exactly, or we have
chosen to make well-motivated changes. For completeness,
we reiterate the salient  features of the SO04  model in our discussion
below.

While in our previous work it was sufficient to track group mass 
evolution to
identify star forming regions, to evaluate the quasar luminosity
function it is necessary to track mergers of groups. We have used the
same method as STD01  for identifying groups, which
relies upon the local baryonic density field to pinpoint centers of mass, and a
spherical overdensity procedure applied to identify the baryon
group. From the baryon group an estimation of the total halo mass is
derived by multiplying by $\Omega_0/\Omega_b$.  The resulting mass
distribution function is in close agreement with that derived from
friends-of-friends, provided suitable limits for the baryon  spherical
overdensity are chosen. To track merger events we rely upon  group
labeling, and we label a merger as having occurred  when at least 30\%
of the accreted mass does not come from a single massive
progenitor. This procedure means that the first groups to form are
also  treated as merger events. For each merger event that will be 
tagged as a quasar we store the details including position and redshift 
in an output file.

Once a group has been identified as satisfying this criterion, the 
dynamical time associated with the cold gas disk which feeds the AGN 
and the mass 
associated with the black hole must be calculated.
For a given redshift, $z$,  
and 
virial density $\rho_{v}(z)$, the implied virial radius for a group of 
$N$ gas particles with mass $m_g$ is
\begin{equation}
r_v=\left[ {N  m_g \Omega_0/\Omega_b \over 4/3 \pi \rho_{v}(z)} 
\right]^{1/3}.
\end{equation}
The circular velocity is then
\begin{equation}
v_c=\left[ {4 \over 3} \pi G \rho_{v}(z) r_v^2 \right]^{1/2},
\end{equation}
and the dynamical time, $t_d$, associated 
with a 
cold disk of gas, is defined by 
\begin{equation}
t_d=0.055\times r_{v}/v_c.
\label{eq:td}
\end{equation}
Note that in order to maintain the same relationship between
outflow velocity  and black hole mass as
Wyithe \& Loeb (2002) and SO04, we choose a slightly larger
time than was used in these studies.  This is because our numerical
model also includes thermal energy input to establish the correct
initial post-shock temperature according to eq.\ (\ref{eq:Ts}) below.

While the observed $M_{\rm bh}-\sigma_c$ relation
(Merrit \& Ferrarese 2001; Tremaine \etal 2002) 
infers that the black hole mass 
scales as $\sigma_c^\alpha$, where $\alpha \sim$ 4-4.5, 
the $M_{\rm bh}- v_c$ has a slightly steeper slope
because the $v_c-\sigma_c$ is shallower than linear (Ferrarese 2002).
Thus we assume an 
$M_{bh}-v_c$ relationship given by,
\begin{equation}
M_{bh}=2.8\times 10^8 \left({v_c \over 300 \; {\rm km 
\;s^{-1}}}\right)^5. 
\end{equation}
Lastly, the black hole is assumed to shine at its Eddington luminosity 
($1.2 \times 10^{38}$ ergs s$^{-1}$ $\msun^{-1}$) for 
the dynamical time, $t_d$.  Note that these values are slightly
different than in SO04, but the overall relationship between $v_c$
and luminosity is the same.

\subsection{Outflow Implementation}

Each AGN in our simulation is assumed to channel a fixed fraction
fraction, $\epsilon_k$, of its bolometric energy into a kinetic outflow. 
The amount of energy deposited into the 
outflow is then
\begin{equation}
E_k=1.2 \times10^{38}\; \epsilon_k \left({ M_{bh} \over 
\msol}\right) \left({t_d \over 
{\rm s}}\right) \;{\rm ergs}.
\end{equation}
As in SO04, we shall adopt $\epsilon_k = 0.05$ throughout this investigation,
which is consistent with other literature estimates (\eg Furlanetto \& Loeb
2001; Nath \& Roychowdhury 2002).
The majority of mass in the outflow at the resolution we can simulate
will have come from material surrounding the cold gas group. 
Therefore, as in our previous work, we model the expanding outflow as a 
spherical shell outside of the virial radius of the system. While the 
assumption of a
spherical shell is a significant oversimplification, given the bipolar
nature of outflows, it is worth recalling that within the intracluster
medium in galaxy clusters a bipolar outflow will still launch an
ellipsoidal cocoon of shocked intracluster gas (Begelman \& Cioffi
1989). Hence, we place the expanding outflow at a 
radius $2r_{vir}$ and re-arrange all gas below a density threshold of 
$2.5\rho_{vir}$ within this radius, 
but 
outside $r_{vir}$, into an expanding shell. 
The density threshold 
prevents us from redistributing cold gas, which is known to be 
very stable against incoming shocks in SPH simulations. 
Once we have established the amount of mass available 
to create the shell, $M_s$, the velocity of the shell, $v_s$, is 
calculated from,  
\begin{equation}
v_s= 1.13 \left( {E_k \over M_s} -G m_g {N \over N_s} {\Omega_0 
\over \Omega_b} \sum^{N+N_{s}}_{i=N+1} {1 \over r_i} - 
{1 \over r_o}\right)^{1/2}, 
\label{eq:vs}
\end{equation}
where the second term on the right-hand side denotes the potential 
energy subtracted as particles are moved from their initial position to 
the shell. As in STD01, we add an additional rotational velocity 
component to the shell so as to preserve the angular momentum.

Note that the prefactor in eq.\ (\ref{eq:vs}) is less than 
$\sqrt{2}$ as
a fraction of $E_k$ is channeled into establishing the 
correct post-shock temperature in the outflowing gas.
This heating is particularly important 
since it will help determine the fraction of impacted gas that is able to
cool within a Hubble time. Under the assumption that the shell behaves like a 
strong shock, the postshock temperature, $T_s$, is
\begin{equation}
T_s={ 3 \mu m_p v_s^2 \over 16 k_B}={ (13.6 K)v_s^2 \over (1 {\rm km \,
s^{-1}})^2 }.
\label{eq:Ts}
\end{equation}
While the semi-analytic model in SO04 assumes that this heating applies 
to the galaxy as well, here we only heat the material in the outflowing shell. 
Our motivation for this choice is the short cooling time of gas in 
galactic halos and, on a secondary level, the collimated bipolar nature 
of 
the outflows. The radial expansion 
of the shell agrees with analytic predictions (STD01).

Since our resolution is insufficient to provide detailed knowledge of the 
inner structure of galaxies 
we implement star formation on the basis of a merger model. Following 
a major merger we convert 10\% of gas in the galaxy into stars 
particles. While this method is known to be a good model of high 
redshift star formation in low mass halos (STD01) it does not track 
quiescent mode star formation which is the primary mode of 
star formation in the higher mass galaxy population. We emphasize that 
the purpose of this study is to focus on the hydrodynamic evolution of 
the IGM and we 
are not attempting to calculate a luminosity function for galaxies, 
instead  
we use them largely as a tracer population.

\section{Distribution and Evolution of Quasars}

\subsection{Quasar Clustering}

In the following three subsections we study the optical properties
of quasars, as quantified in the rest-frame B-band.
In keeping with our previous investigations, as well as
the observations in Elvis \etal (1994),
we relate the luminosity in this band to the overall
bolometric luminosity by assuming
a fixed ratio of  $L_{\rm Bol} =10.4 L_B$ 
at all luminosities and redshifts.  Fixing
this value also allows for direct comparison with Wyithe 
\& Loeb
(2003).  Finally, the $L_{\rm Bol}$ of the quasar 
associated
with each outflow is simply computed as $L_{\rm Bol} =
E_k \epsilon_k t_d^{-1}.$

We begin by addressing the spatial distribution of quasars, which
serves as a check of our merger-based approach.  To quantify this distribution,
we construct the spatial correlation function of quasars using the
center-of-mass information from the outflow data produced in the
simulation.
In principle, this should be computed accounting for the finite
lifetime of each quasar, according to eq.\ (\ref{eq:td}). In
practice, these times are long enough that such effects
can be ignored for distances $\lesssim c t_d \approx 20 (1+z)^{-1/2}$ comoving
Mpc.  Thus we calculate
the 3-dimensional real space correlation function 
using the simplest estimator,
\be
\xi_{qq}(z,m)_k + 1 = {DD(z,m)_k \over RR(z,m)_k},
\ee
where $DD(z,m)_k$ is the number of pairs with a magnitude greater than
some limit, separated by a comoving
difference corresponding to a bin $k$, and $RR(z,m)_k$
is
the average number of pairs that would be found at a given
separation in a random distribution of points
with an overall density equal to the mean density of observable quasars.

We next adopt a fixed magnitude limit in the B-band of
20.84, to allow for comparisons with observations from
the 2dF quasar redshift survey (Croom \etal 2001; 2002),
which have an overall photometric $b$ band limit of 20.9,
where $B \approx b +0.06$ (Goldschmidt \& Miller 1998).
This can be computed as
\[
B = 5.5 -2.5 {\rm log}_{10}\left(\frac{L_B}{L_{\odot}}\right)
          +5 {\rm log}_{10} \left(\frac{d}{10 {\rm pc}}\right)
\]
\be
+2.5(1-\alpha_\nu){\rm log}_{10}(1+z),
\ee
where $d$ is the comoving distance to the quasar, and $\alpha_\nu =
-0.5$ is the typical slope of the quasar power-law continuum 
(Wyithe \& Loeb 2005).

A detailed examination of short range quasar correlations is given in 
the Sloan binary quasar study of Hennawi
\etal (2006). To compare to their sample we project the redshift space
correlation function within a maximum velocity range of $|\Delta
v|<2000$ km s${}^{-1}$, to give 
\begin{equation}
w_{qq}(R,z)= \frac{1}{s_{\rm max}-s_{\rm min}}
\int^{s_{\rm max}}_{s_{\rm min}} \xi_{qq}(R,s,z)ds.
\end{equation}
Here the integration limits are set by $s_{\rm max}=-s_{\rm min}=2000/aH(z)$
corresponding to the width of the velocity
interval. Note that in determining $w_{qq}(R,z)$, rather than integrating over the
redshift space correlation function we instead use the real space 
version. This is a good
approximation since the defined velocity interval is sufficiently large
to contain most of the velocity distribution in the redshift direction.

\begin{figure}
\centerline{\psfig{figure=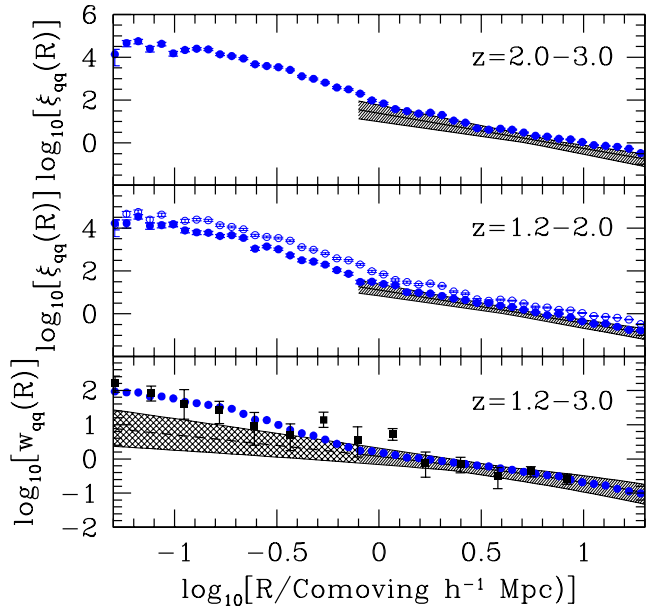,height=9cm}}
\caption{{\em Top:}
Real space correlation function of $B \leq 20.9$ quasars
from $z=2$ to $3$.  The points are taken from
our large 1020 run, while the solid line is the
($2.1 \leq z \leq 2.9$) fit to the 2dF quasar redshift survey
by Croom \etal (2001), bounded by the measurement errors.
Note that this measurement does not extend to separations below
$0.8 {\rm Mpc}$ h$^{-1}.$  {\em Center:} Real
space $B \leq 20.9$ quasar correlation function
from the 1020 simulation from $z=1.2$ to $2$,
as compared with the $1.35 \leq z \leq 1.7$
Croom \etal (2001) results.  Symbols are as in the upper panel,
while the open circles are the simulated $2\leq z \leq 3$
correlation function, shown for comparison.
{\em Bottom:} Angular correlation function of $B \leq 20.9$
quasars.   The solid circles are taken from the simulation at
$z=1.2-3.0$, while the solid line is the Croom \etal (2001)
observations,
projected and averaged over the same redshift range.
The cross-hatched region is an extrapolation of the Croom \etal (2001)
fit to smaller separations, which helps to highlight the
excess in $w_{qq}$ at small separations observed by Hennawi \etal (2005).
This excess is well reproduced in our simulations, where it
arises from gravitationally bound pairs of quasars (the so-called
``one halo contribution'').}
\label{fig:xi}
\end{figure}

In Figure \ref{fig:xi} we plot the real space correlation function in 
the
$z=2.0-3.0$ and $z=1.2-2.0$ ranges. On scales larger than 1 Mpc the
agreement with the Croom \etal (2001) results is strong, with our
results showing a small excess over the mean observational result.
The overlay of the higher redshift results in the middle panel
shows, as expected, that the redshift evolution at these early times is
very weak.
On sub-Mpc scales there is a noticeable turn-up, which is best
examined
using the projected correlation function $w_{qq}(R,z)$, shown in the
bottom panel. To compare to the Croom \etal (2001)
results we have projected their real space correlation function fit, and 
also
extended this fit below their nominal cutoff. This provides a clear
quantitative baseline for examining the short-scale clustering
excess observed by Hennawi \etal (2006).  The precise position and
magnitude of the clustering excess are reproduced
extremely well within our simulation, which we take as both support
for our approach and validation of the Hennawi \etal (2006) results.

This turn up, which has also been observed in the high-redshift
Lyman break galaxy population (Ouchi \etal 2005) as well as
in a local sample of galaxies (Zehavi \etal 2004), is most
likely due to gravitationally bound pairs of quasars that
are orbiting each other.  This so-called ``one halo'' contribution
(\eg Bullock \etal 2002; van den Bosch \etal 2003;
Magliocchetti \& Porciani 2003) should become important at
distances less than
\be
d_{\rm 1halo} \approx
\left[ \frac{m_{\rm Q}}{(4 \pi/3) \Omega_0 \rho_c} \right]^{1/3},
\ee
where $m_Q$ is the mass of the halos associated with the quasars
above our magnitude limit. From the large scale clustering
this is $\approx 2 \times 10^{12} M_\odot$, corresponding to the
$\approx 1.0$ Mpc position of the turn-up, lending further weight
to this interpretation.

\subsection{Quasar-Galaxy Cross-Correlation Function}

As a complementary investigation,
we also examine  the cross-correlation function 
between quasars and galaxies, $\xi_{qg}.$
By cross-correlating these two populations we are directly able to
evaluate whether galaxies containing quasars are clustered differently
than similar-mass quiescent galaxies.
Early observational attempts to measure $\xi_{qg}$ were limited
by sample size and thus a bias toward 2-dimensional angular 
measurements prevailed
(\eg Ellingson, Yee \& Green 1991; Smith, Boyle \& Maddox 1995; Croom \&
Shanks 1999). However, the SDSS and 2dF quasar surveys have enabled
cross correlations in 3-dimensions below a redshift limit of $z<0.3$
(Croom \etal 2003; Wake \etal 2004) and a study at intermediate 
redshifts
using the DEEP2 data has been undertaken (Coil \etal 2006).
To date, these investigations have not uncovered
any bias in $\xi_{qg}$ on scales down to the minimum scale
to which they are sensitive, which is around 1 Mpc.

To evaluate $\xi_{qg}$ we use the quasar catalog from the previous
section, combined with a galaxy catalog evaluated with a  FOF group
finder on the baryonic material in our simulation. We use a linking
length $b=0.065$ to find groups with an outer density
limit of $\delta\simeq 2000$. With a baryonic mass cut of $10^{10.5}$
$M_\odot$ we
find 37,995 groups, and for $10^{11} M_\odot$ we find 11,857 groups.
The estimator for the cross correlation
function is
\begin{equation}
\xi_{qg}(z,m)_k + 1 =
{D_q D_{g}(z,m)_k \over R_q R_g(z,m)_k},
\end{equation}
where $D_q D_{g}(z,m)_k$ is the number of quasar-galaxy pairs
above the magnitude limit in bin $k$, and $R_q R_g(z,m)_k$ is the number
of quasar-galaxy pairs that would be expected if these objects
were randomly distributed with the same densities as in our
simulation.

\begin{figure}
\centerline{\psfig{figure=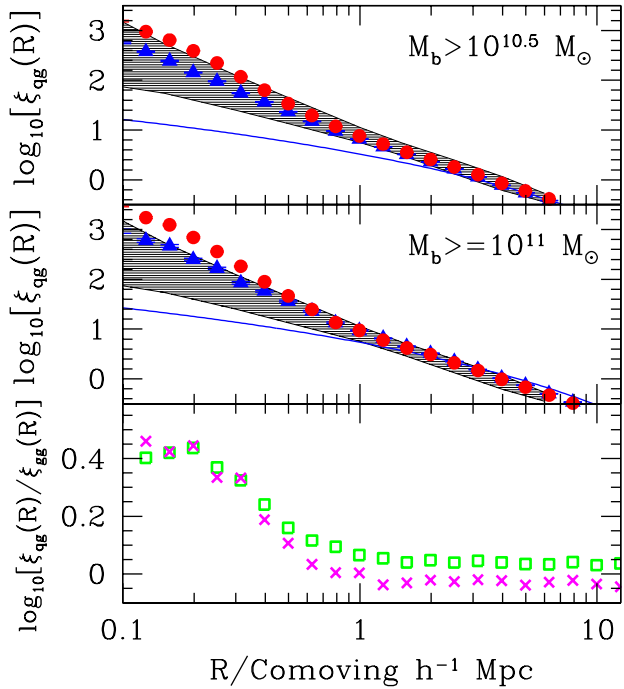,height=10cm}}
\caption{{\em Top:} Real space cross-correlation function, $\xi_{qg}$,
of $M_B \leq -22$ quasars to galaxies
with baryonic masses above $10^{10.5} \msun$ (circles)
as compared to the
autocorrelation function of these galaxies (squares).
The selected mass and magnitude limits have been chosen
to be in broad agreement with the quasars
and galaxies used in measuring the cross-correlation function from
the DEEP2 survey (Coil \etal 2006).  These observations are shown as
the shaded region, which is an estimate that adopts the errors
from the projected cross-correlation function, around the best fit
to the spatial correlation function, $\xi_{qg}(r) = (r/3.45)^{-1.68}.$
Finally the solid lines gives the linear correlation function
of $10^{12} \msun$
halos. {\em Center:} Real space quasar-galaxy cross-correlation function
and galaxy-galaxy correlation function, but now associating
galaxies with objects with baryonic masses above $10^{11} \msun.$  
Symbols
are as in the upper panel.  {\em Bottom:}  Ratio of quasar-galaxy
cross-correlation function to galaxy-galaxy correlation function
in the $M_b \geq 10^{10.5} \msun$ case (squares) and
$M_b \geq 10^{11} \msun$ case (crosses).  The cross correlation
function not only exhibits a break below 1 comoving Mpc $h^{-1},$
but this break is stronger than in the galaxy-galaxy autocorrelation
function.
The strengthening of this break is related to the merger nature of
quasars in our simulation, and is consistent with local measurements 
(Serber
\etal 2006), but unfortunately occurs at too small scales to be measured
at $z \approx 1$ from current surveys.}
\label{fig:xiqg}
\end{figure}

Our results are plotted in Figure \ref{fig:xiqg}, in which we
now employ an {\em absolute} magnitude limit of $M_B = -22,$
to better compare with observations.  On scales 
larger than 1 Mpc, $\xi_{qg}$ is indistinguishable from $\xi_{gg}$, in 
agreement with observations. This is true regardless of whether we 
choose the  $M_b> 10^{10.5}$
$M_\odot$ or  $M_b> 10^{11} M_\odot$ galaxy populations.
However, on scales below 600 h${}^{-1}$ kpc,
$\xi_{qg}$ exhibits a clustering {\em enhancement}.
At first glance, this would appear to be consistent with our results
from \S3.1, which show that one-halo effects can create an excess of
clustering at small scales. However, the explanation cannot be this 
straightforward.
In the bottom panel of Figure \ref{fig:xiqg} we plot the ratio
$\xi_{qg}/\xi_{gg},$ which shows explicitly the turn-over 
from large-scale agreement to short-scale excess. At large 
separations, active galaxies are clustered 
very similarly to the general population, consistent with the DEEP2 results.
Yet at small separations, a dramatic change occurs.
Despite the fact that 
the one-halo contribution is implicitly included in $\xi_{gg}$, {\em 
the 
amplitude of the 
break in the cross correlation function exceeds that of the galaxy 
autocorrelation function by a factor of $\approx 2.5$.}

Thus it appears that our identification of quasars with mergers enhances 
their clustering on the smallest scales. At first it seems that this is 
strongly at odds with previous theoretical studies of mergers (Percival 
\etal 2003; Scannapieco \& Thacker 2003).  However, these studies were 
targeted to separations larger than 1 Mpc, where the two-halo term is 
the dominant contribution to the correlation function.  The excess we 
find here occurs purely in the one-halo regime, meaning that 
mergers have an excess of very close neighbors.

In support of our 
results, local studies of quasar-galaxy clustering have found an excess 
of galaxies at radii $<0.5$ Mpc, completely analogous to the one in our 
simulation.   
Studying a $z < 0.3$ sample of quasars drawn from the
Sloan Digital Sky Survey (SDSS), Serber \etal (2006), have found
that $M_i<-23.3$ quasars are more than three times more clustered
than $L^*$ galaxies on  $\lesssim 0.1$ Mpc $h^{-1}$ scales, although
they cluster similarly to $L^*$ galaxies on larger scales.
While this study was carried out at much lower redshift, these quasars
have roughly the same intrinsic magnitudes as those in Figure 2,
as can be estimated assuming a typical redshift of $z \approx 1.4$, 
which for our simulated sample
gives
a distance modulus of $\approx 45$ or $M_B \lesssim -24.$
Thus there seems to be mounting observational and theoretical
evidence  that the correlation function of the products of mergers, 
while
only very weakly enhanced at large separations, may nevertheless
be more strongly enhanced in the one-halo regime.

In fact, an intriguing possibility is that this is caused by 
three-body interactions in which a third galaxy removes angular
momentum from a nearby close pair.  This suggests that 
dynamical friction may not always be the dominant process driving
galaxy mergers.   Rather, a significant number may be caused
by a process more akin to the formation of tight binaries in
dense star clusters (\eg Rasio, Pfahl \& Rappaport 2000).
Clearly this issue merits future investigation.
 
\subsection{Optical Quasar Luminosity Function}
\label{QLF}

\begin{figure*}
\centerline{\psfig{figure=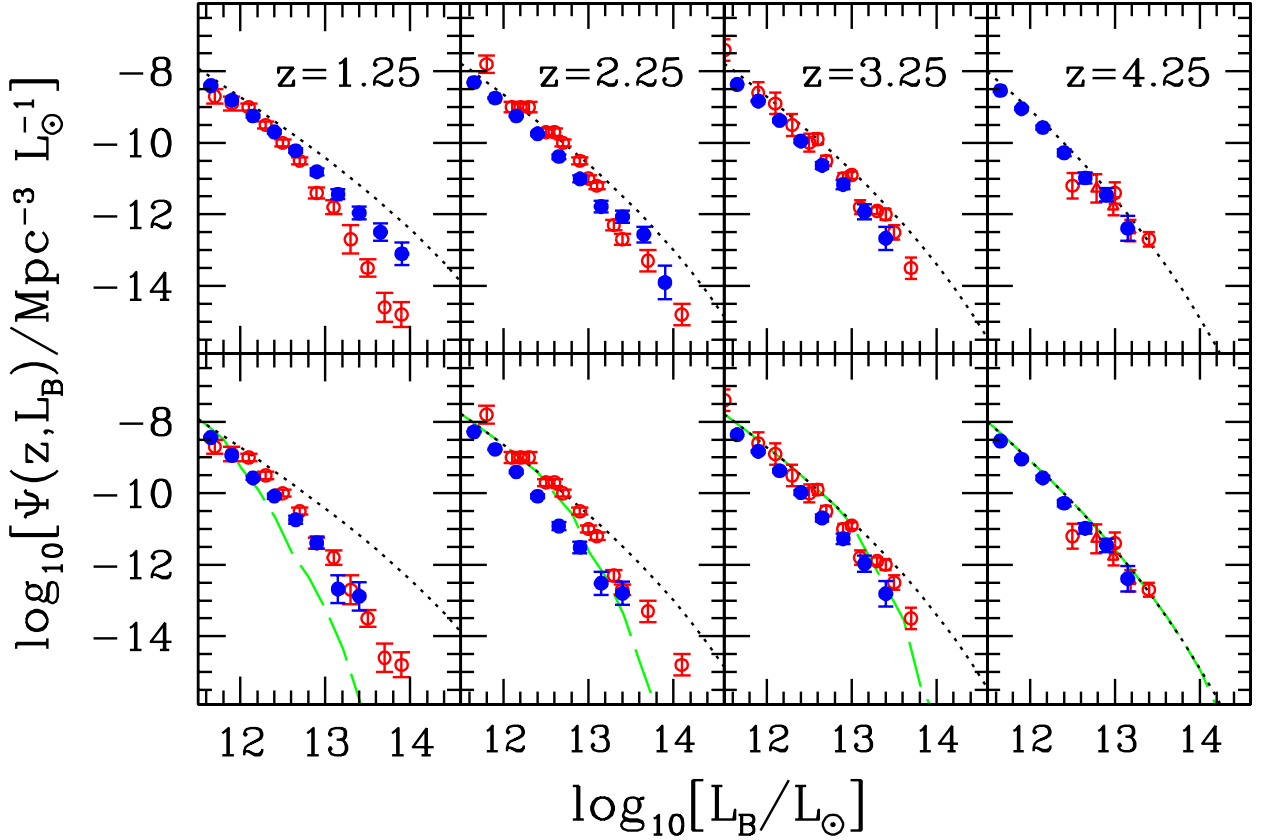,height=12cm}}
\caption{
Evolution of the B-band quasar luminosity function.
Here the data points are taken from
Pei (1995, open circles), which are derived from the compilation by 
Hartwick \& Schade 
(1990), and Fan 
(2001, open triangles).
Simulation results are given by the solid points, while
the dotted line is the simple estimate from the
analytic model by Wyithe \& Loeb (2003), which does not
include feedback.   From left to right the columns
give results at redshifts of 1.2-1.75,
1.75-2.75, 2.75-3.75, and 3.75-4.75 respectively.
The top row shows the raw simulation results, while the 
bottom row
shows the luminosity function derived by applying the same 
physical
model as SO04 to the simulation quasar catalog, along with the fiducial 
SO04 model with a 5\% outflow efficiency (dashed line). The 
difference at low
redshift between the
two models arises because the semi-analytic model treats 
heating
and cooling in
a significantly different way to the simulation.}
\label{fig:lum}
\end{figure*}

To construct the luminosity function for each 
luminosity and redshift bin,
we calculate the number of quasars in this bin
times the total time these objects are shining,
and divide by the time interval, the width of the
bin, and the volume of the simulation.
That is for a given redshift bin $i$ and a given 
luminosity bin
$j$ the luminosity function is simply
\be
\Psi_{i,j} = \frac{1}{V \Delta t_i \Delta L_{B,j}} \sum_{k 
\in {\rm bin_{i,j}}}
t_{d,k},
\ee
where the sum is over all quasars with redshifts and
luminosities associated with the $i$, $j$ bin, which spans 
a time  interval $\Delta t_i$ and a range of luminosities 
$\Delta L_{B,j}.$ In \fig \ref{fig:lum} we plot the resulting
luminosity function
for
our fiducial 1020 simulation.

The dotted line in this plot gives the Wyithe \& Loeb (2003) estimate
of the luminosity function, which is simply based on a merger
prescription and does not account for feedback.  Comparing this
estimate with our simulation results  uncovers a clear turn-down in
the number of $L_B \geq 10^{13} L_\odot$ quasars at $z \lesssim 2$.
However, comparing the simulation results with the measured points
make it immediately clear that this turn down is not as strong as seen
in the observations.  This means that the suppression is much weaker in
the simulation than in the  semi-analytic  SO04 model, which was
found to be a good fit to the observations.
This is a surprising result which needs to be understood 
in more detail, and, to explore this further, we recalculate the 
luminosity function but imposing by hand the precise gas heating 
methodology used in the semi-analytic approach.

As emphasized in Oh \& Benson (2003) the ability of gas to cool is
insightfully described from the perspective of entropy. Under the 
frequently used definition of entropy, $S=T/n^{2/3}$, the isobaric 
cooling time
may be
written,
\begin{equation}
t_{\rm cool}={ 3/2 n k_B T \over n_e^2 \Lambda(T)}=S^{3/2}\left( {3
\mu_e^2 k_B \over 2 \mu^2 T^{1/2} \Lambda(T)} \right) =S^{3/2} F(T),
\end{equation}
where $\mu=0.62$, $\mu_e=1.18$, and 
$F(T)$ serves as temperature dependent normalization.
By equating
the cooling time to the Hubble time, $t_H$, we can derive a critical
entropy value: gas above this entropy limit is unable to cool within the
Hubble time and is thus effectively removed from the galaxy formation
process. The critical value is (SO04),
\[
S_{\rm crit}=(280\; {\rm keV cm}^2)(1+z)^{-1}\left[ {E(z) \over 
(1+z)^{3/2}
}
\right]^{2/3}
\]
\be
\times \left[ {\Lambda(T_{\rm min}) \over 6.3 \times 10^{-22} {\rm \;
ergs \; s^{-1} \; cm^3}} \right],
\end{equation}
where $\Lambda(T_{\rm min})$ corresponds to the minimum cooling time
below $10^8 $ K ($T_{\rm min}\simeq 2.3\times 10^5$ K), and 
$E(z)=[\Omega_m
(1+z)^3 +\Omega_\Lambda]^{1/2}$.

Having defined the concept of critical entropy, we first estimate the 
total amount
of mass that can be heated above $S_{\rm crit}$ using eq.\ 
(17) of SO04,
\[
M_{\rm ex}(\delta,z,M_{bh}) =  4.6 \times 10^{12} \, {\rm 
M_{\odot}} \,
S_{100,\rm crit}(z)^{-1} \,
\]
\be
\times E_{60} \, \delta_{\rm s}^{-2/3} \left( 1+z \right)^{-2}.
\label{eq:Mex}
\end{equation}
We then search to a radius $R={\rm min}(R_s,R_{heat})$, 
where $R_s$ is
the shell radius versus time given by,
\begin{equation}
R_{\rm s} = 1.7 \, {\rm Mpc} \,  E_{60}^{1/5} \, 
\delta_{\rm s}^{-1/5}
\, (1+z)^{-3/5} \, t_{\rm Gyr}^{2/5},
\end{equation}
and $R_{heat}$ is the radius of the region heated above 
$S_{\rm crit}$ as
calculated in
the SO04 model
\begin{equation}
R_{heat} =  5.6 \, {\rm Mpc}
\, S_{100,\rm
crit}(z)^{-1/3} E_{60}^{1/3}
\delta_{\rm s}^{-5/9}
\left( 1+z \right)^{-5/3}.
\label{eq:Rheat}
\end{equation}
Our assumed $S_{\rm crit}$ is 60 keV cm${}^2$,  corresponding to the
metallicity value of $Z=0.05$ used in our simulation, and we taken an
average post-shock overdensity, $\delta_{\rm s}$, of 20. If a quasar
is found within $R$, we subtract $M_{\rm ex}$ from the gas
mass associated with it, and if  $M_{\rm ex}$ exceeds this mass,
we remove it from our catalog altogether.

The resulting luminosity function is plotted in the second  row of
\fig \ref{fig:lum}. In this case, there is significantly better
agreement between the data (and the semi-analytic model) for this
revised luminosity function.  While this model does not precisely 
reproduce
the strong  knee observed in the SO04 model and the magnitude of the
turn-down at higher luminosities, the
improved agreement is compelling.
It is thus clear that the primary cause of  the
difference between these models, is not their varying  approaches to
modeling quasars themselves, but rather in the  simplified exclusion
conditions as described by eqs.\ (\ref{eq:Mex}) -  (\ref{eq:Rheat}).

While these equations capture most of the salient features  of shock
heating, there are two major effects that they  fail to address.
Firstly, they do not differentiate between material within  galaxies
and material in the intergalactic medium, while in our  simulation we
do not apply the outflow heating to the host galaxy itself.  Secondly,
eqs.\ (\ref{eq:Mex}) - (\ref{eq:Rheat}) assign a single density to all
the gas  associated with an object that is overtaken by an outflow,
whereas heating processes in the simulation are directly affected by
the detailed substructure in  this gas.

To explore the relative impact of these two effects,
we examine the distribution of progenitor halo labels in the 
three largest outflows in our simulation at an epoch of $z=1.3$. These 
systems have baryonic masses ranging from $4.7\times 10^{12}$ $M_\odot$ to 
$5.1\times 10^{12}$ \msol, and the majority of particles within them 
have a single label associated with the previous outflow event. To determine 
the number of particles with dissimilar labels we 
place a sphere at the center of mass of the group and then fit a 
radius (by hand) to enclose the outer boundary of the particles with 
the main group label.  
The typical radius of this outer boundary was 250 kpc. For all systems 
we found a significant amount of substructure as evidenced by all groups 
having at least 18, and typically more than 40, distinct labels 
each with at least 20 particles. The total number of particles with 
labels distinct from the main group was always close to the merger mass 
limit, indicative of a system just about to undergo the outflow event 
associated with the merger. 

This results seems to suggest that both effects may be contributing
to the reduced suppression, as each outflow is associated with 
a large group with a single index 
(that might have been disrupted if in-galaxy heating
was included), which merges with a collection of remnants with many 
indices
(that might not have been accreted if in-shock heating were more efficient).
Ultimately, the stronger agreement for the revised catalog versus that
of the original simulation indicates that both of these issues 
need to be explored further.   On the simulation side this would involve
contrasting the present results with a model that includes more efficient 
ejection 
of gas from galaxies themselves.  Likewise, in the semi-analytics
the efficiency of heating could be parameterized
by adding an  additional parameter to account for in-shock
cooling, although it  would be necessary to conduct detailed
simulations of this process to precisely calibrate this
number. These are significant issues which we will return to in the
discussion.

\begin{figure*}
\centerline{\psfig{figure=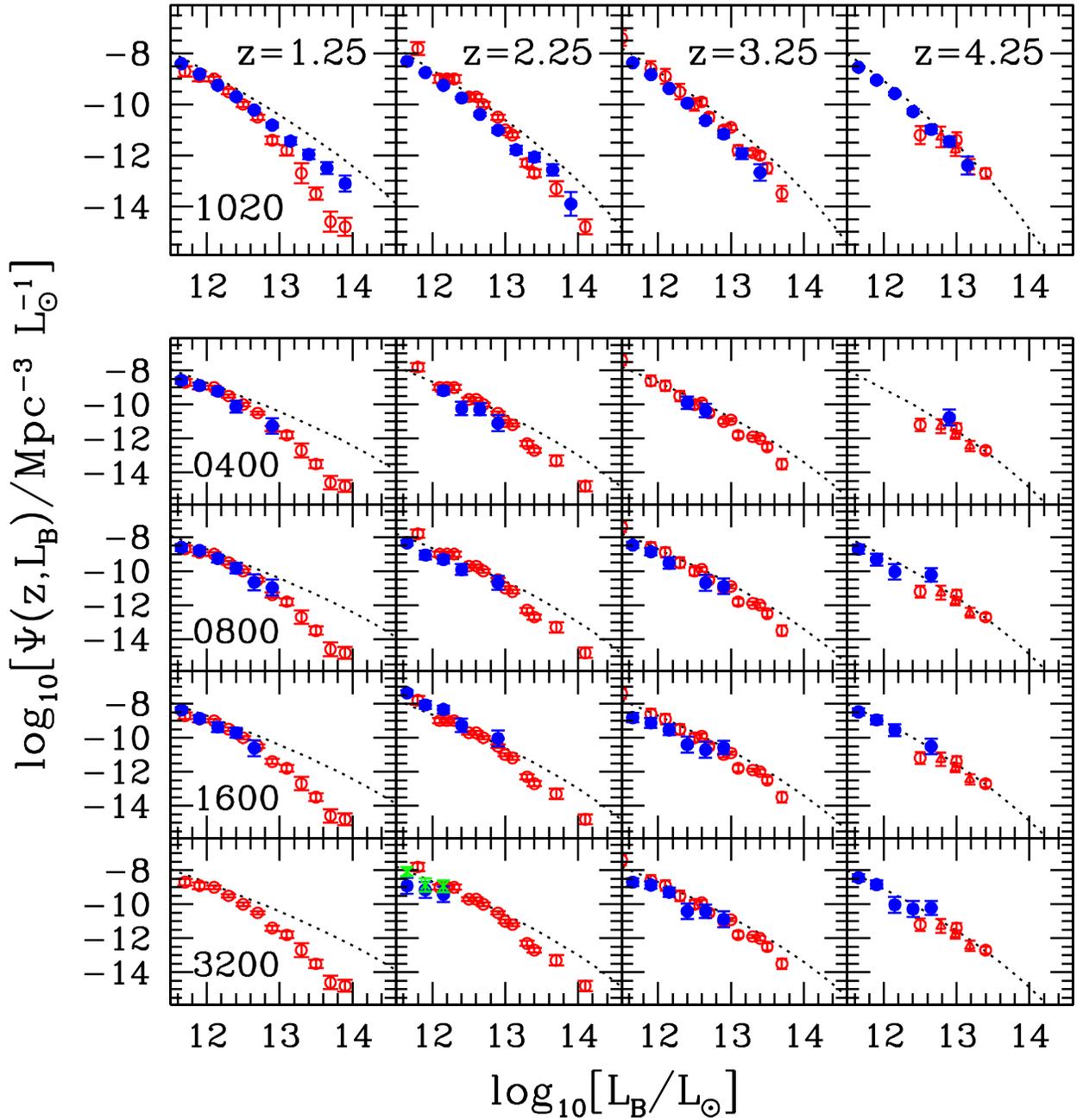,height=19cm}}
\caption{Evolution of the B-band quasar luminosity function
as a function of resolution.  Again
the data points are taken from  Pei (1995, open circles) and Fan (2001,
open triangles),  as compiled by Hartwick \& Schade (1990).
Simulation results are given by the solid points, while the dotted
line is the simple estimate from the  analytic model by Wyithe \& Loeb
(2003), which does not include feedback.    From left to right the
columns give results at redshifts of 1.0-1.75 ($1.2-1.75$ in run
1020),  1.75-2.75 (2.5-2.75 in run 3200),  2.75-3.75, and 3.75-4.75 
respectively.  From top to
bottom, the rows show results from our large ($146 h^{-1}$ Mpc$^3$)
run 1020, and the smaller ($35 h^{-1}$ Mpc$^3$) runs 0400, 0800, 1600,
and 3200.  Symbols are as in Figure \protect\ref{fig:lum}, and, for 
comparison, the  crosses in $z=2.5-2.75$ panel of the 3200 run show the 
results of the 1600 run limited to this same redshift range.}
\label{fig:lum2}
\end{figure*}

Finally, as a test of convergence,
in Figure \ref{fig:lum2} we compare the luminosity function
in our fiducial simulation with that derived from our 
35 $h^{-1}$ Mpc$^3$ simulations.  In this case we do not impose
eqs.\ (\ref{eq:Mex}) - (\ref{eq:Rheat}).   At all redshifts
we obtain good agreement between runs over the luminosity range
spanned by these smaller simulations volumes, although the 400 results
are very noisy.   Note that while
in the $z \approx 2.25$ column, the luminosity function in the 3200 run has 
fewer large quasars than the other runs, this is due to small number statistics
and the fact that it was stopped earlier than the other simulations due 
to the excessively large number of time-steps required.
Similarly, $L_B \geq 10^{13} L_\odot$ quasars are so rare that they
can not be compared between the 1020 simulation and the test simulations, 
motivating our choice of an extremely large volume for this run.

\subsection{Hard X-ray AGN Luminosity Function}

\begin{figure}
\centerline{\psfig{figure=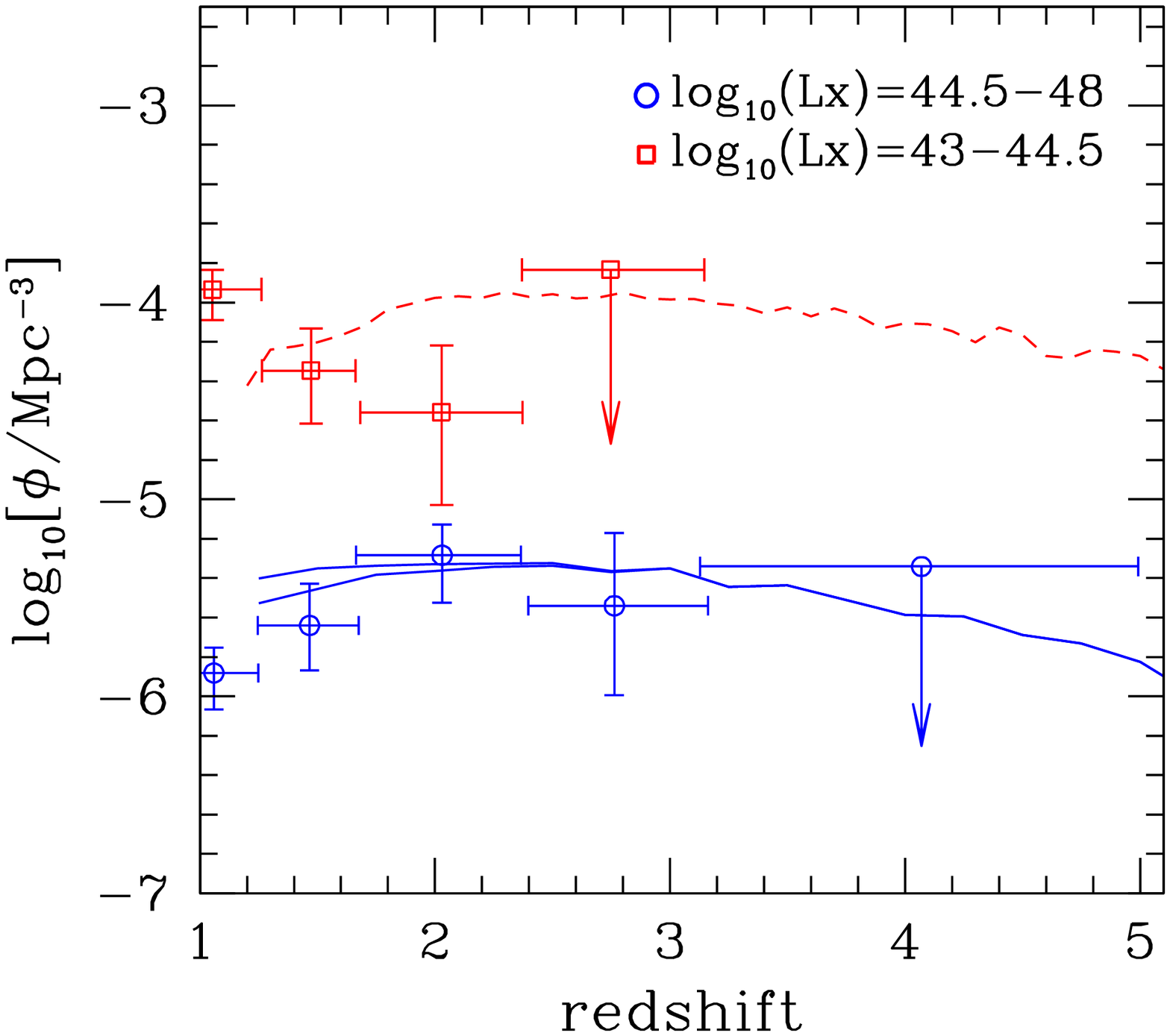,height=9cm}}
\caption{Hard X-ray luminosity function. The solid and dashed
lines give
the number density of X-ray luminous ($44.5 \leq \log_{10}(L_X) \leq 48$)
and X-ray faint ($43 \leq \log_{10}(L_X) \leq 44.5$) AGN
in our 1020 simulation, respectively.
In the luminous case, the lower solid curve shows the results in which
the exclusion conditions, eqs.\ (\ref{eq:Mex}) - (\ref{eq:Rheat}) are imposed.
These models are compared with
observations of luminous (circles) and faint (squares) AGN as
compiled by Ueda \etal (2003).}
\label{fig:ueda}
\end{figure}

Our simulation can also be directly compared with observations of the
hard X-ray luminosity function, which we construct from our
bolometric luminosities using a similar approach to that for the
optical quasar luminosity function. 
To convert from the bolometric luminosity to $L_X$ 
we use
results from Marconi \etal (2004) who calculated the expected
correction for $\approx 10^{12}$ $L_\odot$ AGN with a spectral template
motivated by recent observations. The ratio of the bolometric
luminosity to the hard X-ray band (2-10 keV) is given by a
third-degree polynomial,
\begin{equation}
\log[L/L(2-10\; {\rm keV})]=1.54+0.24{\mathcal L}+0.012{\mathcal
L}^2+0.0015{\mathcal L}^3,
\end{equation}
where ${\mathcal L}=\log(L_{\rm Bol})-12$, and $L_{\rm Bol}$ is given in
$L_\odot$.

Early observational work used ASCA data (Boyle \etal 1998) and BeppoSax data (La Franca
\etal 2002) to show that the hard X-ray luminosity function (HXLF) was
evolving strongly between $z = 0$ and $1.5$, consistent with pure luminosity
evolution. More recently, Cowie \etal (2003)
used {\em Chandra} data in two redshift bins ($z$=0.1-1 and $z$=2-4) to argue
that the AGN number density for luminosities lower than $10^{44}$ erg
s$^{-1}$ seems to peak at a lower redshift than those of higher
luminosity. This antiheirarchical evolution was demonstrated
definitively by Ueda \etal (2003), who carried out a comprehensive
compilation of HEAO1 (Piccinotti \etal 1982; Grossan 1992), ASCA (Ueda
2001, Akiyama \etal 2003) and {\em Chandra} (Brandt \etal 2001) data
to derive the comoving spatial density of AGNs in three luminosity
ranges between ${\rm log}(L_X)=41.5$ and ${\rm log} (L_X)=48$.

In Figure \ref{fig:ueda} we compare our derived spatial densities to the
observational data in the ${\rm log}(L_X)=43-44.5$ and
${\rm log}(L_X)=44.5-48$ luminosity bands in this sample.
Both the qualitative and
quantitative predictions of the simulation agree with the measurements:
the downsizing trend is apparent in both luminosity bands and the
overall normalizations agree well.
However, below $z\simeq 2$, we are faced with two minor issues.
Firstly, 
in the lower luminosity bin it appears that the luminosity function is
turning down slightly too quickly, as the observations suggest that
the turn down in this luminosity range occurs after $z=1$. Secondly,
the brightest band, while turning down at the observed epoch of
$z=2$, does not perfectly follow the observational trend. Imposing
the exclusion conditions represented by eqs.\ (\ref{eq:Mex}) - 
(\ref{eq:Rheat}) improves
this fit somewhat, although these differences are small compared
to the measurement errors from the observations. Imposing
these conditions has no impact on the lower luminosity bin.
In general, these results are consistent with those in
\S \ref{QLF}.

\section{Other Implications}

\subsection{Impact on the Intergalactic Medium}

While our study has been focused on the properties of the quasar
population itself, our simulations naturally have predictions for
the more tenuous gas surrounding large galaxies.
In fact, as discussed above, the most clear observational
evidence for widespread nongravitational heating lies
not in the galaxy population, but rather in the
properties of the diffuse gas in galaxy clusters.

\begin{figure}
\centerline{\psfig{figure=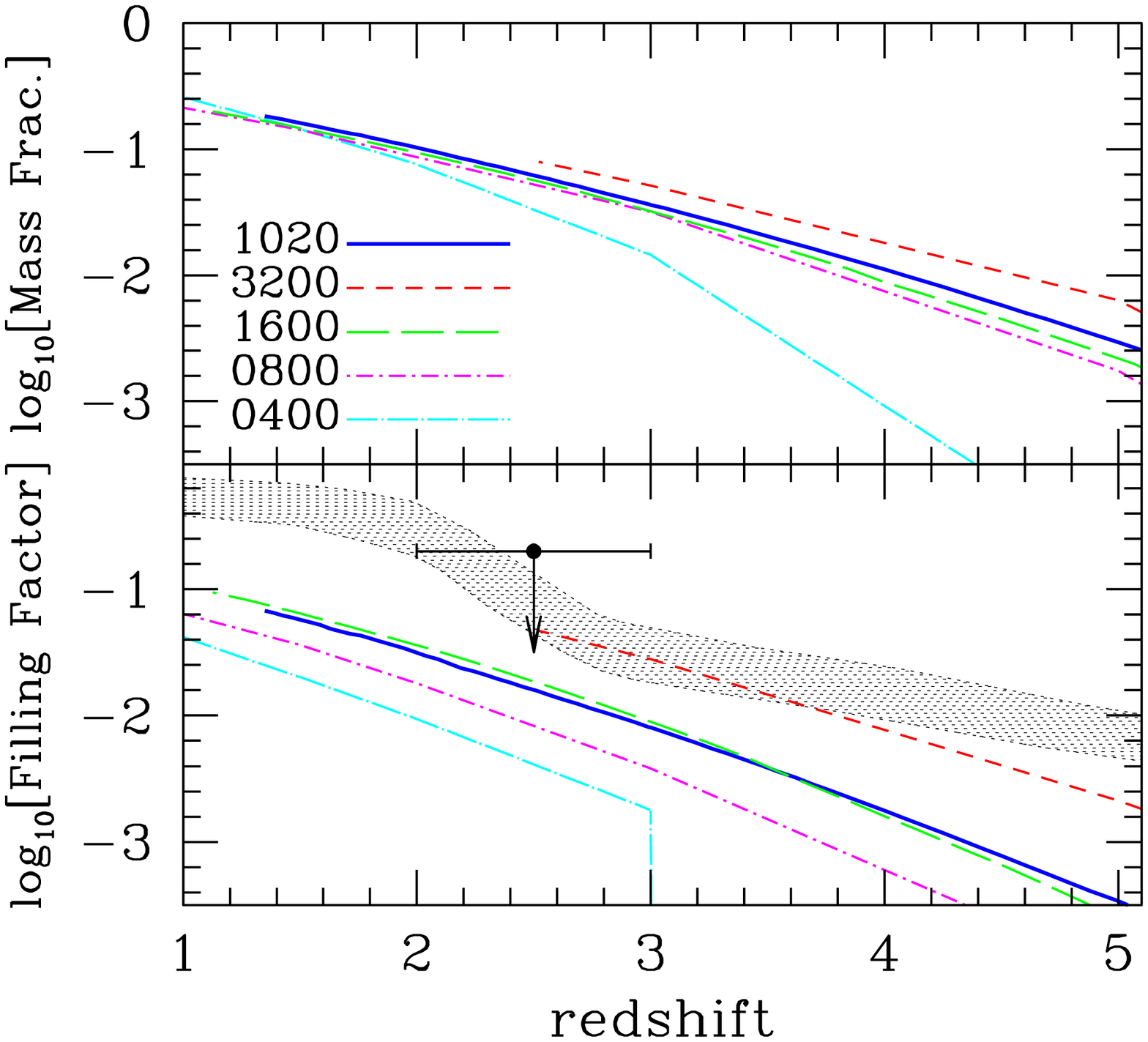,height=9cm}}
\caption{{\em Top:}  Mass fraction of gas heated above $S_{\rm crit}$
as a function of redshift.  The solid line
shows the results from our large 1020 run,  while the
short dashed, long dashed, dot short-dashed, and dot long-dashed, lines
(which move from higher to lower filling factors) show the
results of our 3200, 1600, 0800, and 0400 runs, respectively.
{\em Bottom:}   Volume filling factor of gas heated above
$S_{\rm crit}$ as a function of redshift.  Lines are as in the upper
panel,
while the point gives the Ly$\alpha$ forest constraint discussed in
Levine \& Gnedin (2005, see also Dav\' e \etal 1999).   Finally, the
shaded
region gives the estimate of the volume filling factor impacted by AGN
outflows
in Levine and Gnedin (2005), and is bounded from below by their
$\epsilon_k = 0.05,$ $\tau_{\rm AGN}  =10^7$ yr model and bounded from above
by their $\epsilon_k = 0.05,$ $\tau_{\rm AGN}  =10^8$ yr model.}
\label{fig:FF}
\end{figure}

To examine the impact of outflows on this material we first focus on
the total amount of gas that has been shocked to $S > S_{\rm crit}$,
such that it no longer participates in fueling further generations of
quasars.   The redshift evolution of the mass fraction of this
gas is plotted in
the upper panel of Figure  \ref{fig:FF}, which shows that quasar
feedback is primarily a low-redshift phenomenon.   Thus, above $z
\approx 3$ less than 3\% of the gas in the simulation has been
affected, consistent with the lack of suppression of the luminosity
function at these redshifts.  At lower redshifts, however, the $S >
S_{\rm crit}$ mass fraction grows prodigiously, preventing roughly 20\%
of the gas in the simulation from cooling.   This
is consistent with the turn-down in the luminosity function seen in
Figure \ref{fig:lum}, which  while not as efficiently quenched as
the semi-analytic results, nevertheless differs substantially from the
pure-merger predictions.

As a test of convergence, we also plot in this panel the  $S > S_{\rm
crit}$ mass fraction from each of our smaller  simulations.   These
range from the 0400 run, in which particles are $64$ times more 
massive than in the 1020 simulation, to the 1600 run, 
in which particles are
$0.125$ times the mass of those in the 1020 run.  As increasing 
resolution
adds a
large number of low-mass, high-redshift outflows, the mass fractions
at high redshift increase monotonically with resolution.  At
lower redshift, however, the mass fractions approach each other
asymptotically, and in the important $z \lesssim 3$ range, this
quantity is largely consistent across runs.  However, this mass 
convergence does not give a complete picture of the effect of 
resolution.

In the lower panel of Figure \ref{fig:FF} we plot the evolution  of
the volume  filling factor of $S \geq S_{\rm crit}$ gas in  each of
these runs.  To calculate these quantities in the  SPH method it is
necessary to first smooth the particle data on to a  grid. In the 1020
case we do so on a  $1340^3$ mesh, so that the smoothing scale for the
filling factor at  expansion factor $a$ is $0.155a$ Mpc, which is
considerably above our  minimum smoothing length.   In the other
runs we use a mesh of size twice the particle resolution, for 
example, the $2\times 
160^3$ run was smoothed on to a $320^3$ mesh.

The data point in this panel gives the upper limit of on the volume
filling
factor provided by the Lyman-$\alpha$ forest (Levine \& Gnedin 2005),
which
is well above our results, as expected from the semi-analytic estimates
in SO04.  For comparison, we also plot the results from a range of models 
taken from
the N-body + semianalytic study of Levine \& Gnedin (2005).  While this
is substantially higher than our results, this is to some degree due to
the fact that they computed  the full volume impacted by quasars
outflows, rather than only the volume heated above $S_{\rm crit}$.

\begin{figure}
\centerline{\psfig{figure=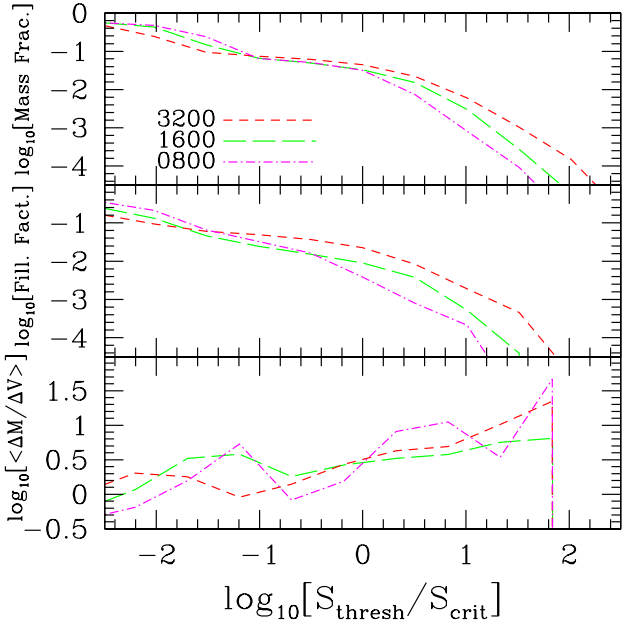,height=9cm}}
\caption{{\em Top:}  Mass fraction of gas heated above a given entropy
threshold, $S_{\rm thresh}$ in or 3200 (short-dashed),
1600 (long-dashed), and 0800 (dot-dashed)
simulations, at a fixed redshift of $z=3$.  {\em Center:}
Volume filling factor above a given threshold at $z=3$.  Lines
as in the upper panel.  For our choice of metallicity,
$S_{\rm crit}(z=3) = 19$ keV cm$^2$, and near this critical value,
the volume filling factor is a strong function of the threshold entropy,
but the mass fraction is almost constant.
{\em Bottom:} Average differential overdensity
(change in mass fraction over change in volume filling factor)
  as a function of $S_{\rm thresh}$.}
\label{fig:FF2}
\end{figure}

Furthermore, it is clear from this figure that the volume filling
factors are  significantly different across simulations, even at the
lowest redshifts. This result seems at odds with our mass-fraction
measurements.  To explore this issue further, in Figure \ref{fig:FF2}
we choose a fixed redshift of $z=3$ and plot the mass fraction and
volume filling fraction above a threshold entropy $S_{\rm thresh},$
which we allow to vary.  For all three models, the mass fraction is
only a weak function of $S_{\rm thresh}$ for all entropy values near
$S_{\rm crit}.$  This means that small differences in entropy  have
only a small impact on the number of particles prevented from cooling,
and therefore both the $S > S_{\rm crit}$ mass fraction (shown in
Figure \ref{fig:FF}) and the luminosity function (shown in Figure
\ref{fig:lum2}) are similar across runs.  Essentially, at $z=3,$ the
particles are divided into two types, those whose entropies are well
above $S_{\rm crit}$, and those that are far below this critical value.

In the lower panel of Figure \ref{fig:lum2}, we plot the volume
filling factor as a function of $S_{\rm thresh},$ again at $z=3.$ In
this case, near $S_{\rm crit}$ the volume filling factor is a strong
function of our threshold entropy.  This suggests that the high
entropy gas is largely found in low density environments, so that
changes in $S$ around $S_{\rm crit}$ pass through a region where the
volume can change rapidly, but there is  actually little mass to
modify the overall mass fraction.

In the lower panel of Figure \ref{fig:lum2} we plot the the
differential overdensity, $\Delta M / \Delta V$, that is  the change
in the mass fraction over the change in the volume filling factor, as
a function of $S$.  This confirms that the majority of $S \approx
S_{\rm crit}$ gas is in environments only a few times denser than the
mean, and that the density of this material is increasing
strongly as a function of entropy.

\begin{figure*}
\centerline{\psfig{figure=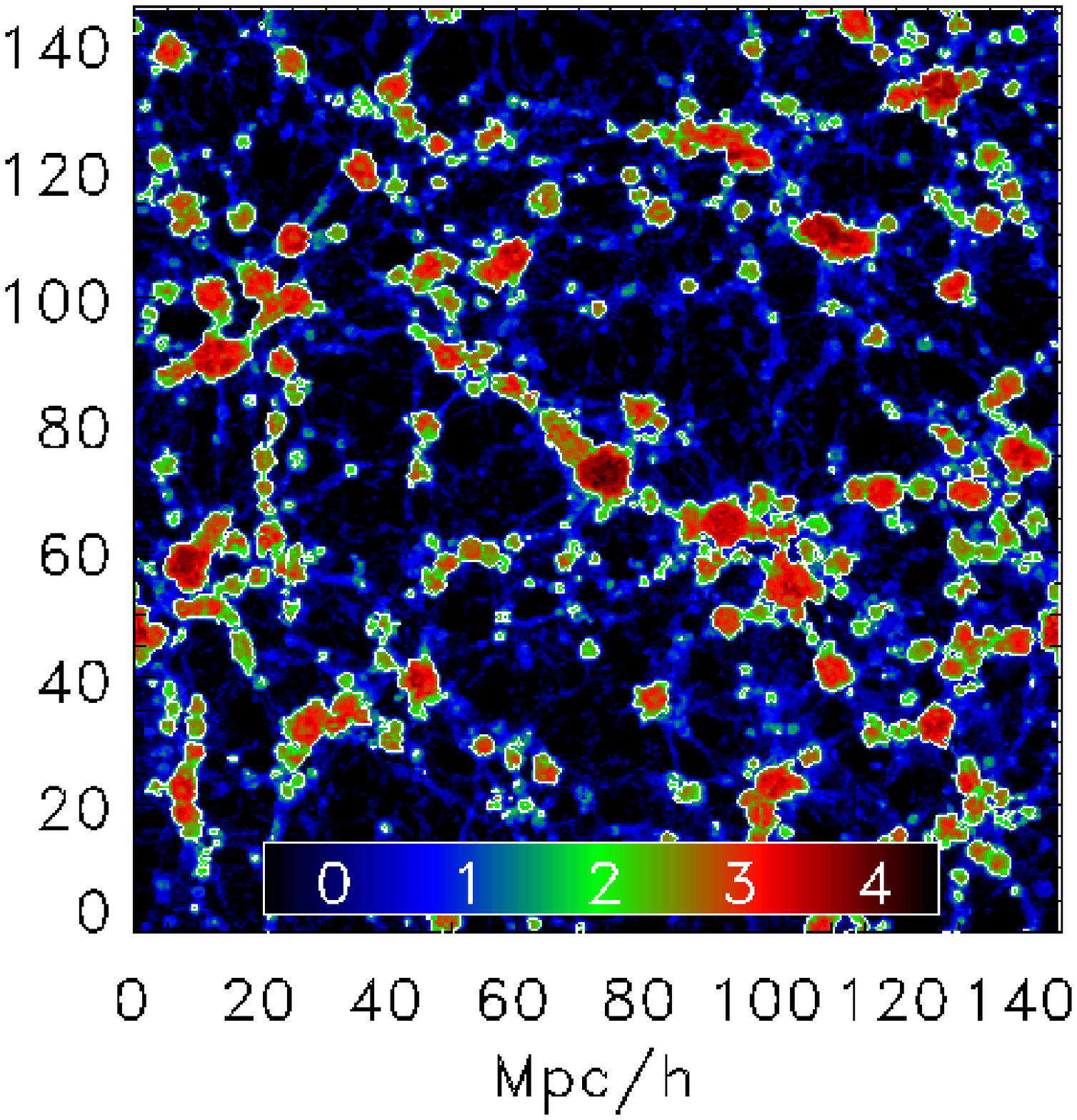,height=8cm}\psfig{figure=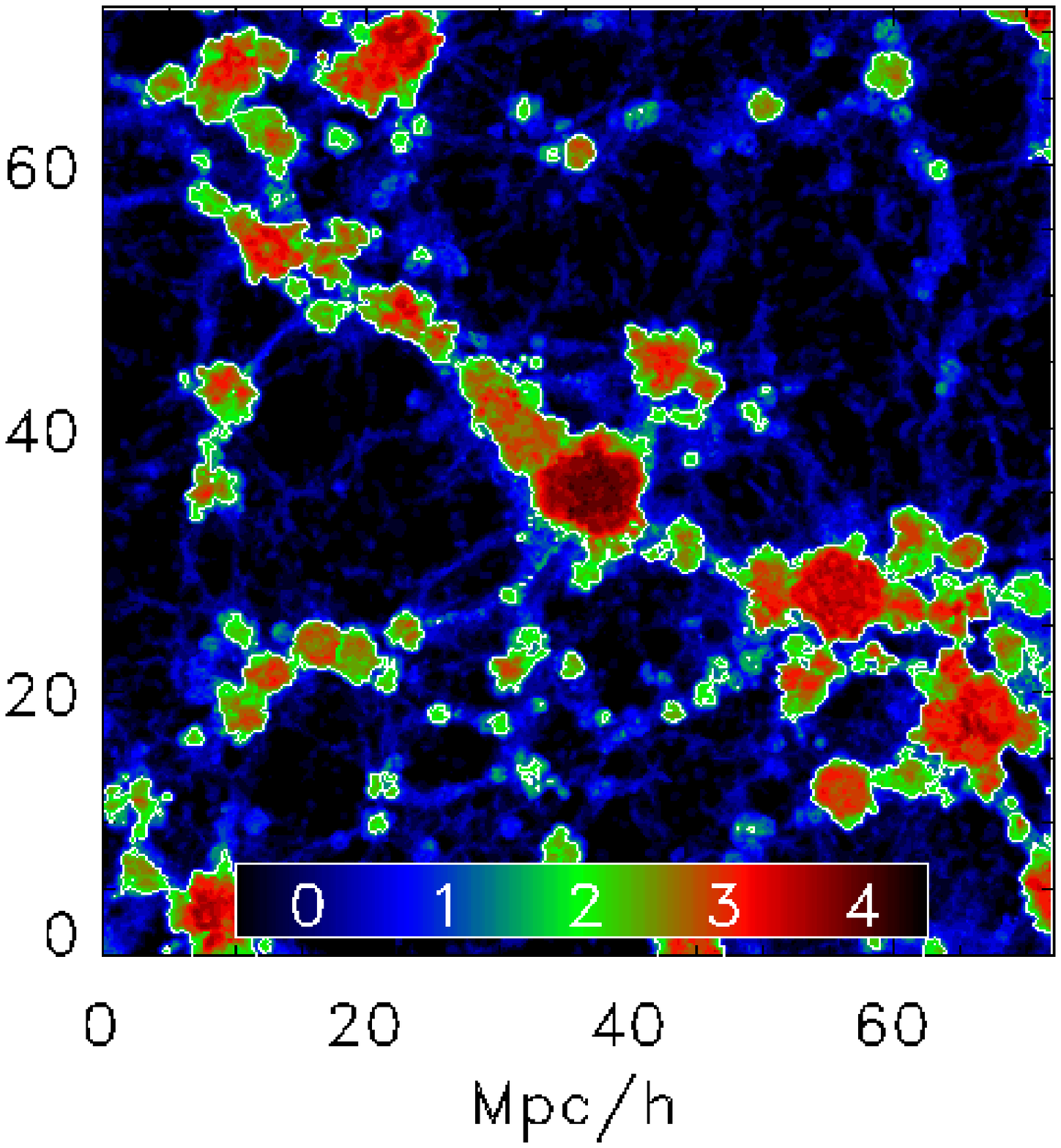,height=8cm}
}
\centerline{\psfig{figure=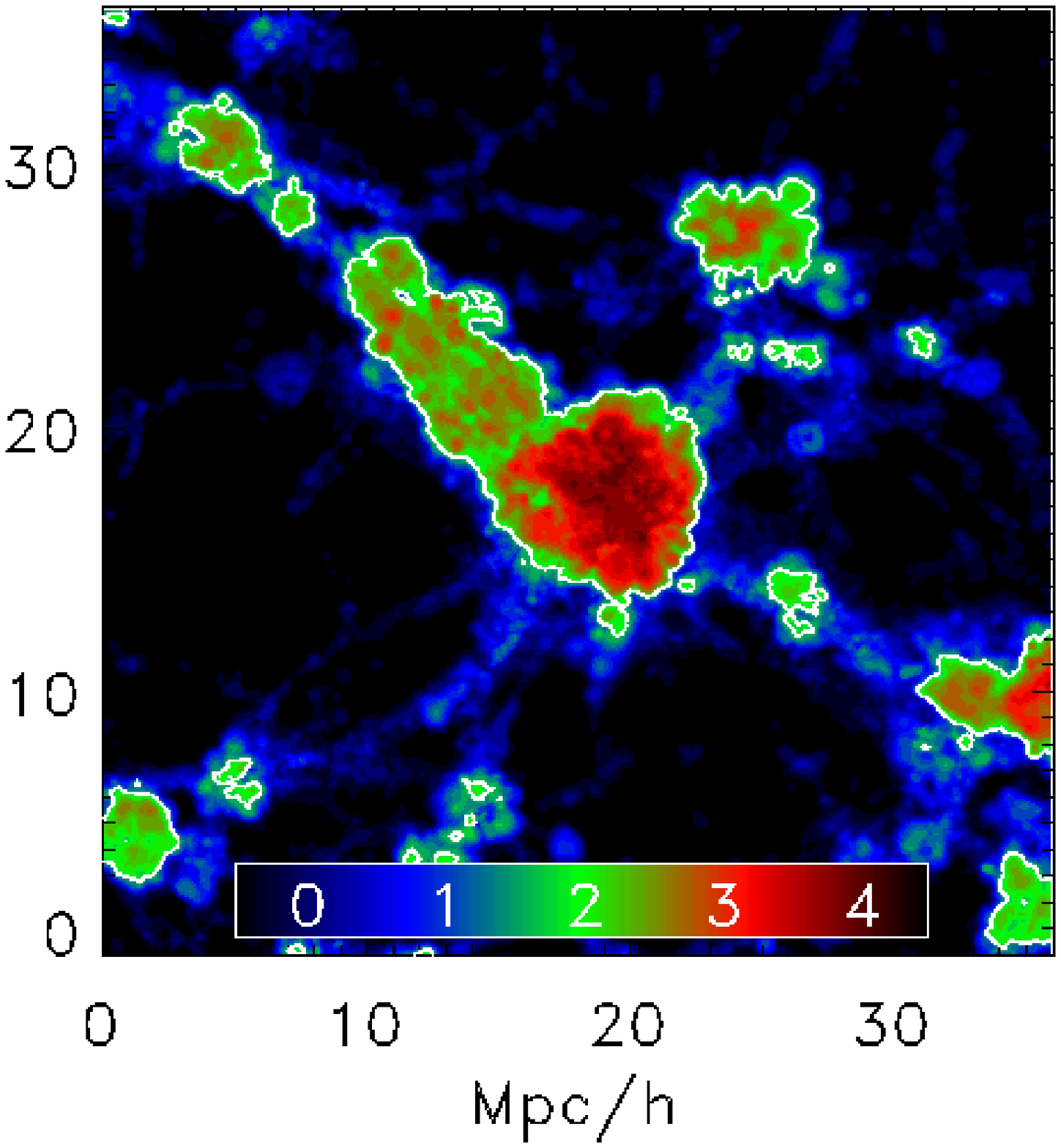,height=8cm}\psfig{figure=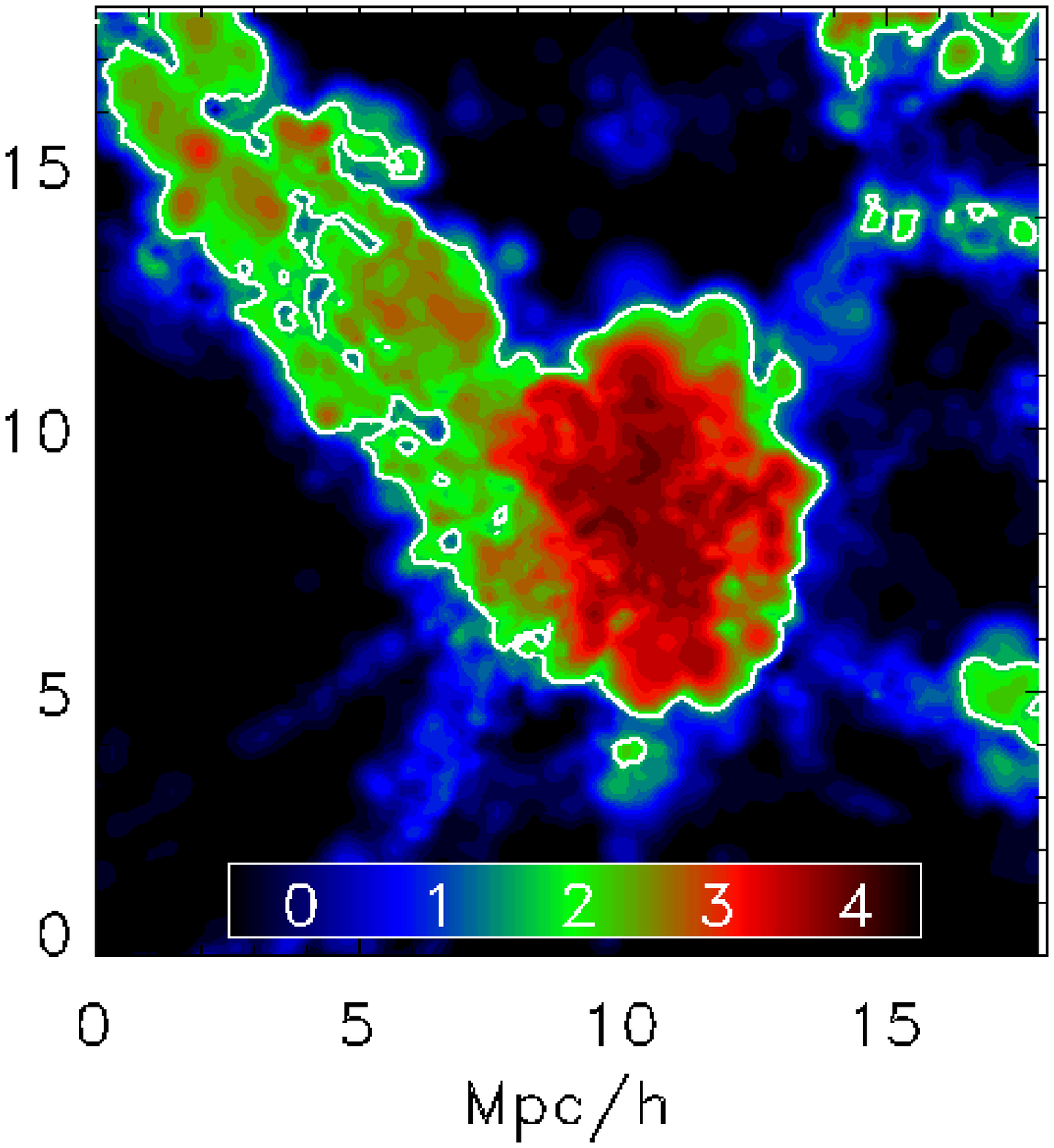,height=8cm}
}
\caption{{\em Top:} Contours of entropy from $z=1.2$ slices through in
our 1020 simulation.  In the upper two panels the thickness of the
slice is 2.4 Mpc $h^{-1}$, while, to emphasize substructure, the
thickness in the 36 Mpc $h^{-1}$ and 18 Mpc $h^{-1}$ panels is taken
to be 1.2 Mpc $h^{-1}$ and $0.6$ Mpc $h^{-1},$ respectively.  All
units are comoving. In each panel the white lines demarcate the
boundary at which $S = S_{\rm crit}.$  In general, this occurs at the
very edge  of halos, where the density is dropping rapidly, but it
also occurs in isolated subclumps in otherwise $S \geq S_{\rm crit}$
regions.}
\label{fig:S}
\end{figure*}
                                                                                
Figure \ref{fig:S} illustrates why this is this case.   Here we show
the entropy distribution in slices taken
from our 1020 simulation at the final output redshift.  It is clear from
this plot that the boundary between $S>S_{\rm crit}$ and subcritical gas
(indicated by the white lines) lies at the very edges of cosmological
halos, where the density is dropping off strongly.  These boundaries are
defined by comparatively few SPH particles, and thus their positions can
change rapidly following small changes in the particle distribution.
Finally, the thin slice shown in the smallest-scale panel in Figure
\ref{fig:S} uncovers the presence of $S \leq S_{\rm crit}$ subclumps
within a larger (high entropy) region.  Again,
the presence of this cooling substructure was not included in the SO04
models, and is one of the key causes of the differences between these
semi-analytical results and those presented here.

\subsection{Impact on Clusters and Groups}

Recent {\em Chandra} observations of Hydra A (Nulsen et al 2005) 
have uncovered evidence for AGN activity within clusters at 
intermediate redshifts. While cooling flow clusters at lower redshift 
seem inconsistent with the idea of powerful shocks (\eg Voit \& 
Donahue 2005, Croton \etal 2006), 
it remains 
an intriguing possibility that these systems underwent an earlier period of 
strong feedback  and have now settled into a quiescent state. 
The observations of radio quiet clusters by Donahue \etal (2005) that 
show extremely long cooling times despite an absence of inferred black 
hole activity, are 
broadly consistent with this hypothesis. These results prompted 
the numerical investigations of Sijacki 
\& Springel (2006) who showed that their model of AGN activity has 
comparatively 
little effect on the cluster $L_X-T$ relationship when all material 
within the virial radius is included. Therefore in this 
section we 
examine the effect of our outflow model on the $L_X-T$ relationship and 
the cluster entropy profile. It is worth noting that as our raw 
luminosity function shows we have an excess of bright quasars at z=1.25, 
the results we derive in this section are an upper bound on the effect 
of outflows on clusters. 

To find cluster groups we first begin from a friends-of-friends b=0.2 
catalogue of the 1020 simulation. The centers of mass are evaluated for 
these groups, and then 
used as the beginning stage of an iterative spherical-overdensity group 
finder that searches radially outward until the group is below the density threshold.
The center of mass is then evaluated and used as the beginning point for the radial 
search, with the process being repeated five times.
This technique has the advantage of biasing against mergers since the center of 
mass of mergers is usually sufficiently offset from the merging groups 
to stop the spherical-overdensity convergence process early, and the group is then 
discarded due to the low amount of mass found. We find 1272 groups by this process.

To estimate the bolometric luminosity of the simulated clusters 
we use
\begin{equation}
L_{\rm Bol}=  
\sum_{i=1}^{N_{group}} {m_i \rho_i \over (\mu m_p)^2} \Lambda(T_i), 
\end{equation}
and the emission weighted temperature is given by,
\begin{equation}
T_{ew}={\sum_i m_i \rho_i \Lambda(T_i) T_i \over \sum_i m_i \rho_i
\Lambda(T_i)}.
\end{equation}
In these formulae, 
$\Lambda(T_i)$
 is the pure bremsstralung estimator of Pearce 
\etal (2000) (see also Navarro \etal 1995 and Muanwong \etal 
2001, for similar approaches), and $m_i, \rho_i, T_i$ are the mass, density and 
temperature of particle 
$i$. While we could use the emission curve 
associated with the $Z=0.05$ metallicity gas used in the simulation, 
the strong peak caused by collisional excitation of He${}^+$ at $10^5$ K 
will give a very high weighting to the cores of clusters, which in fact 
would probably have cooled significantly if we were using a $Z=0.3$ 
metallicity gas. Overall, as emphasized in Muanwong \etal (2001), our 
choice of a pure bremsstralung estimator will bias our luminosities low.

\begin{figure*}
\centerline{\psfig{figure=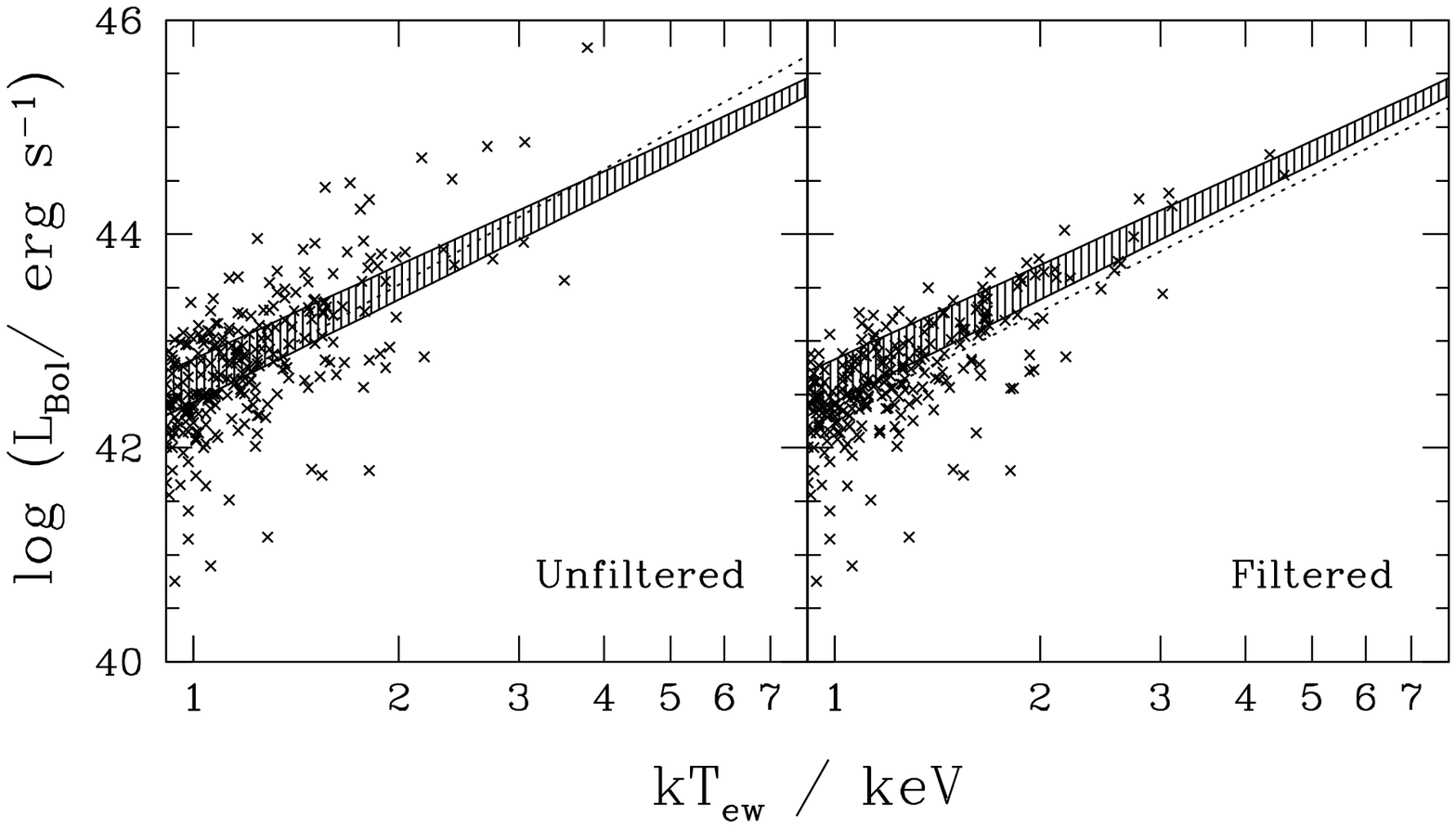,height=9cm}}
\caption{$L_X-T$ relationship for the simulated clusters. On the left 
is the raw unfiltered data including the overly bright core regions 
(with overdensities exceeding $5\times10^4$), 
while the right panel is the filtered data with these regions 
removed. Dotted lines correspond to least squares fits for clusters 
above 1 keV. The shaded 
area corresponds to Maughan 
\etal (2006) fit to the WARPS sample, with the upper limit set by 
taking the maximum normalization, and the lower limit by matching the 
exponent of our best-fit to the filtered data.}
\label{fig:lxt}
\end{figure*}

Under the assumption of self-similar evolution for spherically symmetric 
clusters, and also that bremsstralung emission dominates in the X-ray 
band, the luminosity-temperature relationship scales with redshift 
according to (Kaiser 1986, Maughan \etal 2006),
\begin{equation}
 E(z)^{-1} \Delta(z)^{-1/2} L \propto kT^2f_g,
\end{equation}
where $E(z)$ is the cosmological evolution factor, $\Delta$ corresponds 
to the virial overdensity at a given redshift (\eg see Bryan \& Norman 
1998) and $f_g$ is the cluster 
gas fraction (which is assumed to be independent of $z$ and 
$kT$ following observations of high redshift clusters, \eg Allen \etal 
2002). Thus the redshift evolution in the $L_X-T$ relationship can be 
scaled 
by dividing the luminosity by $E(z) (\Delta(z)/\Delta(0))^{-1/2}$. While 
we could thus scale the $L_X-T$ relationship for $z=0$ clusters  
back to our final redshift, a recent analysis by Maughan \etal (2006) of 
the WARPS clusters in the region $0.6<z<1.0$,
provides an unscaled $L_X-T$ relationship at $z\approx1$. Thus in 
what 
follows we use their results as a comparison. We also note that there 
are no observations of galaxy groups at these epochs.

We plot our results for the $L_X-T$ relationship in Figure 
\ref{fig:lxt}, and also show the fit of 
Maughan \etal (2006). It is immediately noticeable that our raw data 
show a very large scatter in luminosities. Since the luminosity of 
clusters is linearly weighted by the density, we decided to plot  
radial profiles of our clusters to determine whether any ``overcooling" 
effects (\eg Thacker \etal 2000) might be present (see figure 
\ref{fig:clus_pro}). The radial profile shows that the most massive 
clusters do indeed show a sudden upturn in density in their cores along 
the lines observed in test problems. We therefore applied a density-cut 
(filter) at $\delta<5\times10^4$ to the gas in our clusters, which cuts 
out this problem region. The resulting data is plotted in the right hand 
panel of figure \ref{fig:lxt}, and shows a much smaller scatter. 
A least 
squares fit for our 1 keV and brighter unfiltered cluster  
catalogue 
gives 
\begin{equation}
L_{\rm Bol}=2.82\times 10^{42} \left( { T_{ew} \over {\rm keV}} 
\right)^{3.58},
\end{equation}
while the filtered catalogue gives,
\begin{equation}
L_{\rm Bol}=2.09\times 10^{42} \left( { T_{ew} \over {\rm keV}}
\right)^{3.18},
\end{equation}
which is closer to, albeit slightly less luminous than, the Maughan 
\etal (2006) best fit of $5.4\times10^{42} 
( { T_{ew} / {\rm keV}})^{2.92}$. Overall, these results are 
strongly supportive of the conclusions of Sijacki \& Springel (2006). 

\begin{figure}[t]
\centerline{\psfig{figure=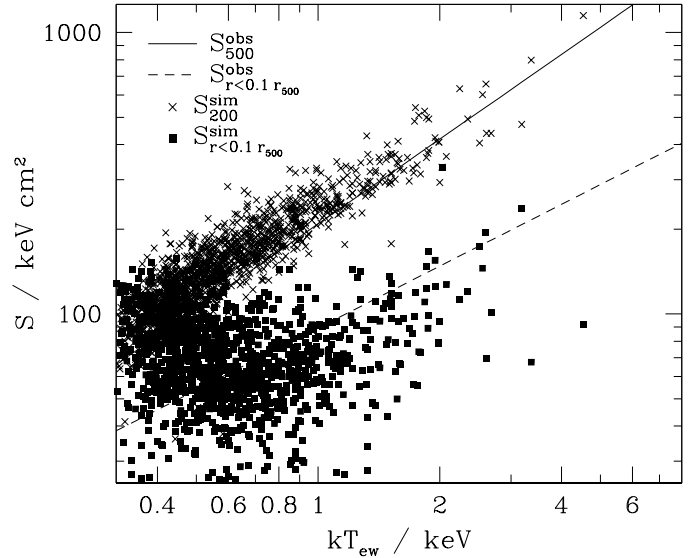,height=8cm}}
\caption{Entropy versus emission weighted temperature for different 
radial cuts. The 4-pointed crosses correspond to the entropy within the 
virial radius and are compared to the ($z\approx 0$) self-similar 
($S\propto T$) fit of Ponman 
\etal 
(2003)  
for the entropy within the $\bar{\delta}_{500}$ region, with the 
normalization taken to match the eight hottest clusters in their 
sample. 
The solid 
squares denote the entropy within 0.1$\,r_{200}$, 
a 
measure of the cluster core entropy. The dashed line is the Finoguenov 
\etal (2005) weighted orthogonal regression fit to the Ponman \etal 
(2003) results. Our results show a significant turn-up in the core 
entropy at groups scales ($<1$ keV).}
\label{fig:clus_ent}
\end{figure}

Lower redshift observations of groups (\eg Xue \& Wu 2000), show the power law 
exponent for groups is close to 5. For systems below 1 keV (regardless 
of whether or not they are core filtered) we do not observe any 
steepening of 
the $L_X-T$ relationship, and there is evidence of the relationship 
becoming shallower, indicative of the gas in this systems being quite 
strongly perturbed. Thus, rather than analyzing relaxed systems, we an 
are in fact analyzing groups that are radiating significantly due to 
the presence of an outgoing shock. In the event that this shock is
sufficient to heat gas above $S_{\rm crit}$, we might well expect, in 
the
absence of significant further accretion and AGN activity, that these
groups expand and cool before $z=0$. To quantify this hypothesis, 
we plot the cluster and core entropy versus temperature 
in Figure \ref{fig:clus_ent}. Our results are in broad agreement with 
the analysis of nearby clusters (Ponman \etal 2003, Finoguenov \etal 
2005), and do indeed show that groups tend to show an excess of entropy 
in their cores.  
Therefore, we tentatively suggest 
that the X-ray emission of galaxy groups at $z>1$ may well be a 
``smoking gun'' for the AGN heating hypothesis. 

\begin{figure*}[t]
\centerline{\psfig{figure=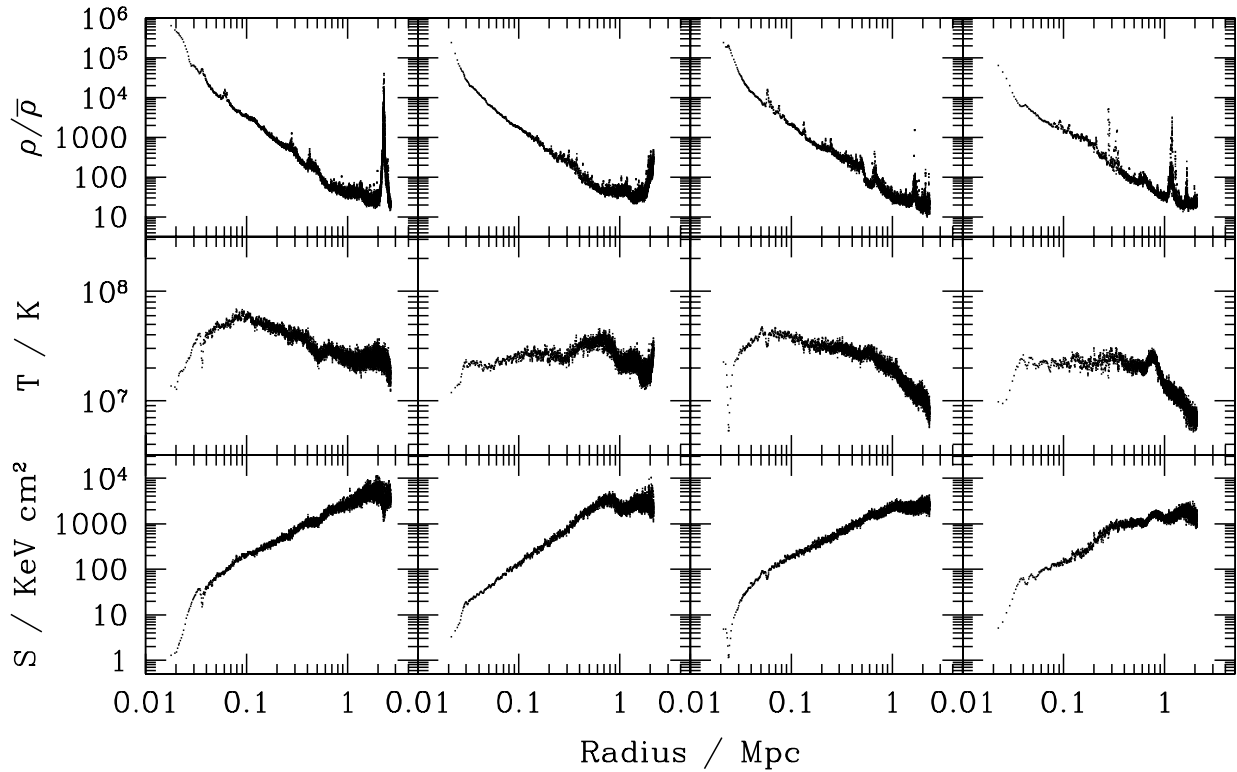,height=11cm}}
\caption{Density, temperature and entropy profiles for the four largest 
relaxed clusters at $z=1.2$ in physical Mpc. A 100 particle moving 
average has been used to smooth 
the data. }
\label{fig:clus_pro}
\end{figure*}

As well as the radial density profiles, we also show the  temperature 
and 
entropy profiles in Figure
\ref{fig:clus_pro}.
The only unusual feature in the temperature profiles is that the second
cluster from the left has a very slightly inverted temperature profile
following an extremely strong outflow event. The resulting entropy
profile is $S\propto r^{1.7}$, while the remaining profiles all match
the $S\propto r^{1.1}$ profile reported elsewhere (\eg Voit, Bryan \&
Kay 2005).

\section{Discussion \& Conclusions}

We have presented results from a suite of cosmological simulations that
self-consistently follow the evolution of quasars and the outflows
associated with them. By tracking the merger history of halos and
applying the quasar model of Wyithe \& Loeb (2003), we have
been able to make direct predictions for the spatial distribution and
luminosity function of these objects. Our results
are in excellent agreement with the observed correlation function of
quasars on both small and large scales, reproducing both the power-law
behavior measured by the 2DF redshift survey on $\geq$ 1 Mpc scales, 
and the strong break measured from the SDSS on $\leq$ 1 Mpc scales.

Furthermore, we predict that the quasar-galaxy cross-correlation
function should show a small scale up-turn relative to the
galaxy-galaxy correlation function.  This occurs on
sub Mpc scales, and therefore can only be measured from
a very large sample of quasars, making its detection 
difficult even in the large DEEP2 survey. However, it is  worth
noting that the {\em Dark  Energy Survey} will produce a quasar
catalog spanning 5,000 square  degrees which combined with photometric
redshifts for 300 million galaxies should be able to suppress
statistical uncertainties to a sufficiently low level to measure this
upturn. We also note that this clustering excess is in  qualitative
agreement with the increase in the integrated galaxy  overdensity at
small scales for the SDSS quasar sample (Serber \etal  2006). Although the
origin of the excess is uncertain, there is an
interesting possibility that the presence of an ancillary galaxy 
can accelerate the merger process via a three-body interaction.
Given the extended nature of the mass distribution associated with
galaxies it is not clear whether this mechanism will work in a
similar way to stellar interactions. We plan to investigate this
intriguing idea in the near future.

As the correlation function is dominated by more populous
low-luminosity quasars, our spatial results should be largely
interpreted as lending support to our dark-matter modeling choice of
merger model.  Similarly, as it tracks the number of quasars formed as
a function of the gas mass accreted  onto galaxies,  the luminosity
function is more sensitive to the details of our feedback modeling.
In this case, our simulation qualitatively reproduces the observed
anti-hierarchical  behavior, but
the turn-down is much weaker than observed.
Matching the suppression at the brighter end requires that we increase
the efficiency  of heating from the AGN outflows to mimic that assumed
in semi-analytic models. This is equivalent to suppressing one key
physical process, namely in-shock cooling in the presence of 
substructure, and including the ejection of gas from quasar host 
galaxies.  
These results
emphasize how sensitive the luminosity function is to issues in the
baryon physics and how the treatment of these issues in semi-analytic
models is still quite approximate.

Investigation of the mass fractions and filling factors of gas above
$S_{\rm crit}$ showed good convergence in the mass fractions at low
redshift, but less so in the filling factors. The differences between
runs are most noticeable at high redshift, where successively higher
resolution leads to the first generation of AGN outflows occurring at
earlier epochs. While mass fractions show fairly strong convergence
below $z\approx 2$, the tendency of gas above $S_{\rm crit}$ to occupy 
low
overdensity regions makes an accurate calculation of the volume
filling  factor difficult due to sampling issues. These results are
also further  complicated by the known dependence of shock resolution
on particle  number (\eg Thacker \etal 2000) where accurate modeling
of shock jumps  in  spherical collapse is reached once
$N_{collapse}>30,000$.  Nonetheless, the results do elucidate that the
impact of  quasar outflows is largely felt at low redshifts, in
agreement with the downsizing trend. Perhaps the most interesting
issue we have not yet explored with regards to 
filling factors is the relative impact of including a more bi-polar
outflow. However, there is no reason to expect  a difference beyond a
factor of two as the gas will flow toward low
overdensity regions, as observed in supernova outflow calculations (STD01).

In somewhat denser environments, our results for the cluster $L_X-T$ 
relation are in broad agreement with observations. This is especially 
encouraging as our cluster modeling is not as sophisticated as other 
more targeted studies.  In this case, the predominant view is that AGN 
heating is best modeled in terms of ``hot bubble'' ejection from the 
brightest cluster galaxy (BCG) which then mixes with the surrounding 
material.  Interestingly, the exact nature of this mixing is not well 
understood, and substantial differences are found between Eulerian and 
Lagrangian simulations, which by definition differ significantly in 
their treatment of advection.  While SPH simulations inhibit the 
development of instabilities that promote mixing due to the necessary 
use of flow interpenetration suppression, it has the advantage of 
exhibiting significantly less numerical diffusion than many Eulerian 
methods. Ultimately, input from laboratory experiments may well be 
necessary to help determine the correct mixing behavior. We also note 
that due to our inefficient star formation model, our simulations 
include a significant amount of cold gas in BCGs that will tend to 
promote an ``overcooling" instability, which others have avoided by 
applying phase decoupling (Pearce \etal 2000).  We believe that this is 
a significant contributor to the lack of an entropy core in our radial 
profiles, and removing this central core produces an $L_X-T$ 
relationship that is significantly less noisy while not exhibiting a 
large change in normalization at intermediate cluster mass scales.  
Additionally, the lack of a strong turn-down in the quasar luminosity 
function at $z=1.2$ also promotes extremely strong outflows in these 
clusters that may well be transporting high entropy gas from the core 
out to the edges of the cluster without significantly raising the 
entropy of the gas immediately surrounding the BCG.

Perhaps the most intriguing result to come out of our $L_X-T$ study is
the prediction that the gas in $z \approx 1$ galaxy groups 
should be strongly perturbed by AGN activity, which is in the process of turning-off 
at that epoch.   This effect is confined to small and high-redshift 
clusters for two main reasons.
At lower redshifts, the comparative paucity of AGN
activity will allow group gas to evolve largely adiabatically, decreasing
$L_X$ dramatically to 
establish the steep $L_X-T$ relation ($L_X \propto T^5$) observed locally.
In more massive $z \approx 1$ clusters, on the other hand, 
the effect of outflows is largely masked when
averaging over the material within $r_{200}.$  Thus 
galaxy groups at $z \approx 1$ 
represent the key mass and redshift range at which the AGN heating
hypothesis is most likely to be testable though observations.

Lastly, our simulation results indicate that simplifications in
current semi-analytic models may well be downplaying critical
physics. While much attention has been paid to drawing broad
conclusions from comparisons between observations and these models,
the differences we have uncovered in this investigation are
troubling.  While, as expected, post processing of our simulation
results was able to reproduce the semi-analytic behavior, it is clear
that in-shock cooling due to substructure in low-overdensity halos is
an issue that must be considered carefully in semi-analytic
calculations.   Fortunately,  constraining the effect by simulation
would not be difficult,  and ultimately a single parameter could be
used to quantify this behavior. On the simulation side, it also is
clear that a more detailed understanding of gas ejection from quasar
hosts will be necessary for definitive conclusions to be reached.
Nonetheless, the insight gained from this initial simulation study
clearly serves to highlight the promise of a self-regulating picture of
quasar formation.  The simple merger  model of Wyithe \& Loeb (2003),
when supplemented with an outflow model, appears to represent a
significant step forward in understanding the evolution of quasar
clustering and the cause of their antiheirarchical low-redshift
turn-off.

\acknowledgments

We thank Andrea Ferrara, Piero Madau and Mattias Steinmetz for hosting
the  IGM-Galaxy Interactions workshop at the KITP where part of this
work was undertaken.  We are also grateful to Alison Coil, Darren Croton,
Marc Davis, Joseph Hennawi \& Lars Bildsten for their many useful
comments, particularly in regards to our comparisons with observations.
Computing was performed at WestGrid (under a RAC
grant), SHARCNET and  HPCVL. We thank Rob  Simmonds of WestGrid for
helping arrange a period of  dedicated use of their IBM p595 system
during initial  testing.  R.J.T.\ acknowledges funding from the CITA
National  Fellow program and NSERC of Canada via the  operating grants
of Professors Larry Widrow and Richard Henriksen.  E.S.\ was supported
by the National Science Foundation under  grants PHY99-07949 and
AST02-05956.

\fontsize{10}{10pt}\selectfont

\end{document}